\documentclass{lmcs}

\usepackage{amsmath, amssymb, graphicx, algorithm, algpseudocode, mathtools, float}

\usepackage{amsthm}
\usepackage{tikz}
\usetikzlibrary{arrows,automata,decorations.pathreplacing}
\usepackage{hyperref}

\newtheorem{example}{Example}
\newtheorem{definition}{Definition}

\title{Decentralized Runtime Verification for LTL Properties Using Global Clock}

\begin{document}


\author[M.~A.~Dorosty]{M. Ali Dorosty}
\address{College of Engineering, Department of Electrical and Computer Engineering, University of Tehran, Iran}
\email{ma.dorosty@ut.ac.ir}

\author[F.~Faghih]{Fathiyeh Faghih}
\address{College of Engineering, Department of Electrical and Computer Engineering, University of Tehran, Iran}
\email{f.faghih@ut.ac.ir}

\author[Ehsan~Khamespanah]{Ehsan Khamespanah}
\address{College of Engineering, Department of Electrical and Computer Engineering, University of Tehran, Iran; Computer Science Department, Reykjavík University, Iceland}
\email{e.khamespanah@ut.ac.ir; ehsahk@ru.is}

\begin{abstract}
  \noindent Runtime verification is the process of verifying critical behavioral properties in big complex systems, where formal verification is not possible due to state space explosion.
  There have been several attempts to design efficient algorithms for runtime verification.
  Most of these algorithms have a formally defined correctness property as a reference and check whether the system consistently meets the demands of the property or it fails to satisfy the property at some point in runtime.
  LTL is a commonly used language for defining these kinds of properties and is also the language of focus in this paper.
  One of the main target systems for runtime verification are distributed systems, where the system consists of a number of processes connecting to each other using asynchronous message passing.
  There are two approaches for runtime verification in distributed systems.
  The first one consists of centralized algorithms, where all processes send their events to a specific decision-making process, which keeps track of all the events to evaluate the specified property.
  The second approach consists of distributed algorithms, where processes check the specified property collaboratively.
  Centralized algorithms are simple, but usually involve sending a large number of messages to the decision-making process.
  They also suffer from the problem of single point of failure, as well as high traffic loads towards one process.
  Distributed algorithms, on the other hand, are usually more complicated, but once implemented, offer more efficiency.
  In this paper, we focus on a class of asynchronous distributed systems, where each process can change its own local state at any arbitrary time and completely independent of others, while all processes share a global clock.
  We propose a sound and complete algorithm for decentralized runtime verification of LTL properties in these systems.
\end{abstract}

\maketitle

\section{Introduction}
Reliability is one of the main characteristics of new complex software used in safety-critical systems.
Many algorithms and solutions have been proposed in order to ensure this requirement.
A popular category of system analysis methods, known as \textit{model checking}~\cite{modelChecking} involve offline observation of the system's model and checking for possibilities of violation of the desired property.
These methods are often expensive and suffer from the well-known state space explosion problem due to the exponential growth nature of the state space in these kinds of methods.
Also, we don't always have access to the underlying system's model.
\textit{Runtime verification}~\cite{rv,rv2} is another category of system analysis methods, where instead of detecting the possibility of property satisfactions/violations beforehand, they are detected them at runtime.
These methods are especially useful when formal verification methods are not able to analyze all the possible executions of the system.
Shortly speaking, runtime verification is applicable when the goal is to analyze a complex system for which model checking is not possible due to the state space explosion, but the property is relatively small, and can be checked using lightweight methods at runtime.

Most of these methods take a formally defined property, describing the correcct behaviour of the system, as input in order to monitor the system at runtime with respect to the given property.
The goal is to determine whether the system's behavior would satisfy or violate the property.
A commonly used language for defining these kinds of properties is LTL~\cite{ltlCtl}, which is the language we focus on in this paper.
Our algorithm is designed for a class of systems known as \textit{distributed systems}\cite{distributedSystems}, which usually consist of a set of processes using message passing as their primary communication mechanism.
The popularity of these types of systems has led to demand for various algorithms defined in a distributed context, including runtime verification algorithms.
Runtime verification algorithms typically take a property $\phi$ as input, which is the property to be checked.
In distributed systems, a common approach to solving this problem is the \textit{centralized} approach, where all processes send their data to a pre-selected process (also known as the \textit{central monitor}) that is responsible for deciding about satisfaction or violation of the property $\phi$.
In order to reach this decision, other processes need to inform the decision-making process about all of their local events that may result in the satisfaction or violation of $\phi$.
This needs a high number of messages to be sent to one process, which results in congestion in that process's incoming channels, as well as making that process a single-point-of-failure.
In the other approach, known as \textit{distributed algorithms}, processes work collaboratively towards a common goal; until one of them deducts and broadcasts the final result of the algorithm.

A number of algorithms have been proposed for distributed runtime verification. Some of them are centralized or semi-centralized algorithms~\cite{centrMTL,centrMTLFreeze}, and hence, suffer from all the previously stated problems. Some define their own property definition languages~\cite{PT-DTL,DTL} (which are less expressive than LTL), and some require the system to be synchronous~\cite{sync,sync2,sync3}.
Falcone et al. in~\cite{falcone} proposed a decentralized runtime verification algorithm that utilizes the global clock, similar to our proposed algorithm. However, our algorithm is much more efficient and needs significantly fewer messages compared to their approach.

In this paper, we propose a decentralized algorithm for runtime verification of LTL properties.
In order to utilize our method, each process must be augmented with the same exact monitoring algorithm, the details of which will be discussed in Sect.~\ref{sect:algorithm}.
This algorithm, which we will refer to as the \textit{monitor} from now on, constantly runs in parallel with the process's main task until it detects the satisfaction or violation of the property.
It is expected that a process's monitor can access its entire local state atomically.
All monitors take the same LTL property as input, at the systems initialization time, and start running at the same time as their corresponding processes.
Each monitor listens for when its process experiences a change in its local state in a way that may result in the satisfaction/violation of the given LTL property.
Monitors are expected to be able to send and receive messages to each other.
Furthermore, all monitors share a global clock, meaning that messages can be timestamped, and the concept of \textbf{``time $t$"} has a common meaning among different monitors.
Our algorithm is \textit{sound} and \textit{complete}, meaning that if one monitor reports the satisfaction/violation of $\phi$ at time $t$, then $\phi$ is actually satisfied/violated at $t$, and vice versa.
In some cases multiple monitors may arrive at the same verdict and report it together.

The rest of the paper is organized as follows: In Sect.~\ref{sect:preliminaries} some preliminary definitions will be given, along with some examples to provide a sense of how we approach runtime verification. We then present our algorithm, first expressing the problem in Sect.~\ref{sect:problem}, and then describing the runtime verification algorithm in full detail in Sect.~\ref{sect:algorithm}, followed by a proof of its soundness and completeness in Sect.~\ref{sect:proof}. We then demonstrate our algorithm's usefulness by showing some experimental results in Sect.~\ref{sect:experiments}. In Sect.~\ref{sect:relatedWork} we review some previous work in distributed runtime verification, and discuss the advantages and disadvantages of each approach. Finally, we conclude by talking about the achievements of our work and discussing future  work that can be possible in Sect.~\ref{sect:conclusionFuture}.

\section{Preliminaries}
\label{sect:preliminaries}

In this section, we give a brief introduction to the concepts we use in our algorithm.
We will use a running example throughout the paper to clarify these concepts and the algorithm details.

\begin{example}
  \label{eg:main}
  Say we have 3 autonomous flying drones that communicate via asynchronous message passing named ``drone A", ``drone B" and ``drone C".
  The mission of all drones is getting to a specified destination.
  Drone A is the leader and drones B and C are the followers.
  We want to verify property $\phi$ at runtime which is defined as:
  $$
  \begin{tabular}{|p{0.1\textwidth}p{0.7\textwidth}p{0.1\textwidth}|}
    \hline
    &\ & \\
    &$\phi \equiv$ \textit{At some point, the leader arrives at the destination. It then stays there until both followers have also arrived.}& \\
    &\ & \\ \hline
  \end{tabular}
  $$
\end{example}

In this paper, we focus on a type of distributed systems, where processes communicate via asynchronous message passing; if a process wants to send data to another process, it sends that data via one or many messages to the other process.
Messages can be received arbitrarily later than the time they were sent and in an order different from the one they were sent.
Also, we rely on the existence of a \textit{global clock} that can be accessed by all the monitors. This can be done by either reading the clock value from one physical clock, or synchronizing local clocks of processes periodically, such that their drift at any arbitrary time is negligible. More formally, we define a distributed system as follows:

\begin{definition}[Distributed System]
  A distributed system $S$ consists of a set of processes $p_1, p_2, \dots, p_n$ that are connected via asynchronous message passing channels.
\end{definition}

\begin{example}
  In our running example, the distributed system can be defined as $\{p_1, p_2, p_3\}$, where $p_1$ is drone A, and $p_2$ and $p_3$ are drone B and drone C, respectively.
\end{example}

Atomic propositions are propositional statements intrinsic to the system and are used to specify the desired properties.
The system designer who comes up with the LTL property, has to first define a set of atomic propositions $AP$ in the system that are relevant to the desired property.

Let $AP_i \subset AP$ be the subset of $AP$ containing atomic propositions belonging to $p_i$.
These subsets of $AP$ are mutually disjoint: $i \neq j \Rightarrow AP_i \cap AP_j = \O$

\begin{example}
  In Example~\ref{eg:main}, we could define the following atomic propositions:
  $$
  \begin{tabular}{|p{0.1\textwidth}p{0.7\textwidth}p{0.1\textwidth}|}
    \hline
    &\ & \\
    &$a \equiv$ \textit{Drone A (the leader) is currently at the destination.}& \\
    &\ & \\
    &$b \equiv$ \textit{Drone B is currently at the destination.}& \\
    &\ & \\
    &$c \equiv$ \textit{Drone C is currently at the destination.}& \\
    &\ & \\ \hline
  \end{tabular}
  $$
\end{example}

\begin{definition}[State]
  A state $\sigma$ is a valuation of all atomic propositions in the system ($AP$).
  Each state is depicted with a subset of $AP$, indicting the atomic propositions that are $\mathit{true}$ in that state.
  Any atomic proposition that is not a member of the set is $\mathit{false}$ at that state.
  The set of all possible states of the system is denoted by $\Sigma = 2^{AP}$ (also known as the systems \textit{alphabet}).
  A \textit{local state} of a process $p_i$ is a valuation of the atomic propositions of $p_i$ ($AP_i$).
\end{definition}

\noindent \textbf{Note:} We use the terms \textit{state} and \textit{global state} interchangeably throughout the paper.

Next we move on to defining our property.
LTL properties are defined on \textit{infinite computations}, which are defined as follows:
\begin{definition}[Finite/Infinite Computation]
  A finite/infinite sequence on the system's alphabet is called a finite/infinite computation (execution) of the system.
  For example, $\omega = \{\sigma_0, \sigma_1, \sigma_2, \dots\}$ is an infinite computation, and $\omega' = \{\sigma'_0, \sigma'_1, \sigma'_2, \dots, \sigma'_k\}$ is a finite computation, where for all $i$, $\sigma_i \in \Sigma$.
  The set of all finite computations of a system is represented by $\Sigma^*$ and the set of all infinite computations of a system is represented by $\Sigma^\omega$.
\end{definition}

\begin{example}
  \label{eg:computations}
  Some examples of finite computations in Example~\ref{eg:main} are as follows:

  \begin{center}

  \begin{figure}[H]
  \begin{tikzpicture}[scale=1]
    \draw [thick] (0,0) -- (9,0);
    \draw [thick,->,gray] (9,0) -- (12,0);
    \node[align=center] at (0.6,-1.3) {No drone has\\initially reached\\the destination};
    \draw [thick,decorate,decoration={brace,amplitude=6pt,raise=0pt,mirror}] (-0.25,-0.15) -- (2.05,-0.15);
    \node[align=center] at (3.75,-1.3) {Drone A is at the\\destination from\\$t=2.1$ to $t=5.2$};
    \draw [thick,decorate,decoration={brace,amplitude=6pt,raise=0pt,mirror}] (2.15,-0.15) -- (5.15,-0.15);
    \node[align=center] at (7.1,-1.5) {Drones A and B\\are at the\\destination from\\$t=5.2$ to $t=9$};
    \draw [thick,decorate,decoration={brace,amplitude=6pt,raise=0pt,mirror}] (5.25,-0.15) -- (8.95,-0.15);
    \node[align=center,gray] at (11.1,-1.3) {All drones are at the\\destination from\\$t=9$ onwards};
    \draw [thick,decorate,decoration={brace,amplitude=6pt,raise=0pt,mirror},gray] (9.05,-0.15) -- (12.25,-0.15);
    \node[align=center] at (2.1,2.5) {At $t=2.1$\\drone A\\(the leader)\\arrives at\\the destination};
    \draw [thick,->] (2.1,1.1) -- (2.1,0.15);
    \node[align=center] at (5.25,2.1) {At $t=5.2$\\drone B arrives at\\the destination};
    \draw [thick,->] (5.2,1.3) -- (5.2,0.15);
    \node[align=center] at (9,2.1) {At $t=9$\\drone C arrives at\\the destination};
    \draw [thick,->] (9,1.3) -- (9,0.15);
    \node[align=center,gray] at (11,0.8) {\textbf{Satisfaction}};
    \draw [line width=1.6pt,->,gray] (11,0.5) -- (9.3,0.15);
  \end{tikzpicture}
  \caption{A sample execution scenario resulting in the satisfaction of $\phi$}
  \end{figure}
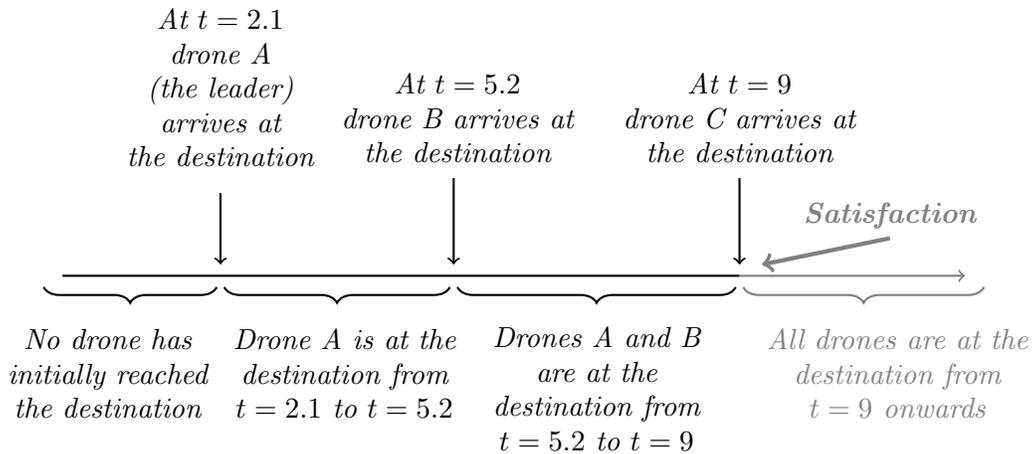

  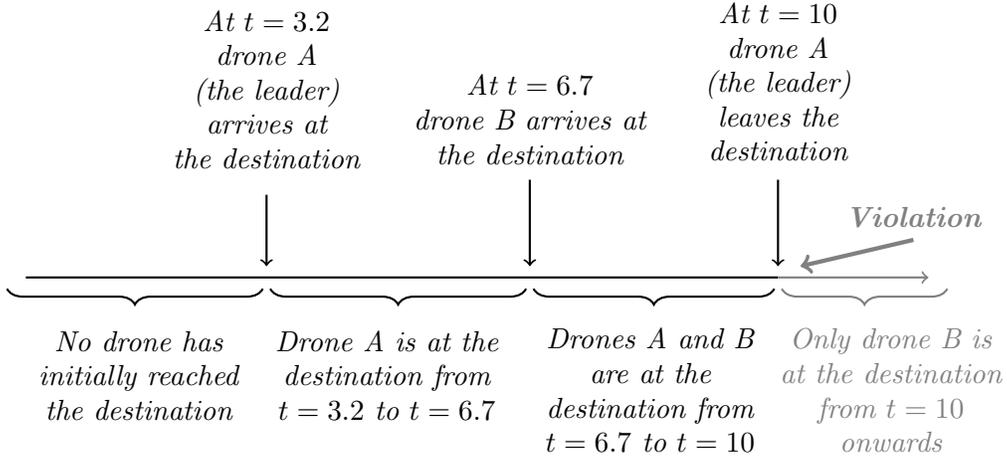
\begin{figure}[H]
  \begin{tikzpicture}[scale=1]
    \draw [thick] (0,0) -- (10,0);
    \draw [thick,->,gray] (10,0) -- (12,0);
    \node[align=center] at (1.5,-1.3) {No drone has\\initially reached\\the destination};
    \draw [thick,decorate,decoration={brace,amplitude=6pt,raise=0pt,mirror}] (-0.25,-0.15) -- (3.15,-0.15);
    \node[align=center] at (4.8,-1.3) {Drone A is at the\\destination from\\$t=3.2$ to $t=6.7$};
    \draw [thick,decorate,decoration={brace,amplitude=6pt,raise=0pt,mirror}] (3.25,-0.15) -- (6.65,-0.15);
    \node[align=center] at (8.3,-1.5) {Drones A and B\\are at the\\destination from\\$t=6.7$ to $t=10$};
    \draw [thick,decorate,decoration={brace,amplitude=6pt,raise=0pt,mirror}] (6.75,-0.15) -- (9.95,-0.15);
    \node[align=center,gray] at (11.5,-1.5) {Only drone B is\\at the destination\\from $t=10$\\onwards};
    \draw [thick,decorate,decoration={brace,amplitude=6pt,raise=0pt,mirror},gray] (10.05,-0.15) -- (12.25,-0.15);
    \node[align=center] at (3.2,2.5) {At $t=3.2$\\drone A\\(the leader)\\arrives at\\the destination};
    \draw [thick,->] (3.2,1.1) -- (3.2,0.15);
    \node[align=center] at (6.7,2.1) {At $t=6.7$\\drone B arrives at\\the destination};
    \draw [thick,->] (6.7,1.3) -- (6.7,0.15);
    \node[align=center] at (10,2.6) {At $t=10$\\drone A\\(the leader)\\leaves the\\destination};
    \draw [thick,->] (10,1.3) -- (10,0.15);
    \node[align=center,gray] at (11.8,0.8) {\textbf{Violation}};
    \draw [line width=1.6pt,->,gray] (11.8,0.5) -- (10.3,0.15);
  \end{tikzpicture}
  \caption{A sample execution scenario resulting in the violation of $\phi$}
  \end{figure}

  \end{center}

\noindent Note that in this example, $a$ can only be checked by drone A, and similarly, $b$ and $c$ can only be checked by drone B and drone C, respectively.
\end{example}

\begin{example}
  \label{eg:computationsWithAP}
  Using states (sets of atomic propositions), the scenarios in Example~\ref{eg:computations} can be shown as:

  \begin{center}

  \begin{figure}[H]
  \begin{tikzpicture}[scale=1]
    \draw [thick] (0,0) -- (9,0);
    \draw [thick,->,gray] (9,0) -- (12,0);
    \node[align=center] at (0.9,-0.7) {$\sigma_0 = \{\}$};
    \draw [thick,decorate,decoration={brace,amplitude=6pt,raise=0pt,mirror}] (-0.25,-0.15) -- (2.05,-0.15);
    \node[align=center] at (3.65,-0.7) {$\sigma_1 = \{a\}$};
    \draw [thick,decorate,decoration={brace,amplitude=6pt,raise=0pt,mirror}] (2.15,-0.15) -- (5.15,-0.15);
    \node[align=center] at (7.1,-0.7) {$\sigma_2 = \{a,b\}$};
    \draw [thick,decorate,decoration={brace,amplitude=6pt,raise=0pt,mirror}] (5.25,-0.15) -- (8.95,-0.15);
    \node[align=center,gray] at (10.65,-0.7) {$\sigma_3 = \{a,b,c\}$};
    \draw [thick,decorate,decoration={brace,amplitude=6pt,raise=0pt,mirror},gray] (9.05,-0.15) -- (12.25,-0.15);
    \node[align=center] at (2.1,1.3) {$t=2.1$};
    \draw [thick,->] (2.1,1) -- (2.1,0.15);
    \node[align=center] at (5.25,1.3) {$t=5.2$};
    \draw [thick,->] (5.2,1) -- (5.2,0.15);
    \node[align=center] at (9,1.3) {$t=9$};
    \draw [thick,->] (9,1) -- (9,0.15);
    \node[align=center,gray] at (11,0.8) {\textbf{Satisfaction}};
    \draw [line width=1.6pt,->,gray] (11,0.5) -- (9.3,0.15);
  \end{tikzpicture}
  \caption{A sample computation ($\sigma = \{\{\}, \{a\}, \{a,b\}, \{a,b,c\}\}$) resulting in the satisfaction of $\phi$}
  \end{figure}
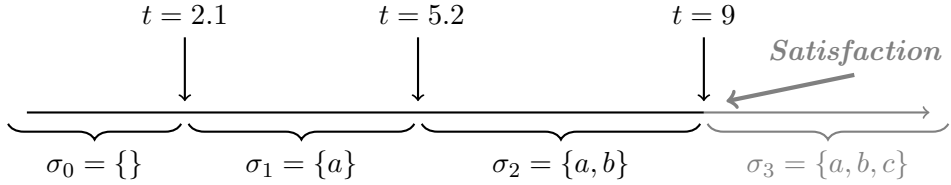

  \begin{figure}[H]
  \begin{tikzpicture}[scale=1]
    \draw [thick] (0,0) -- (10,0);
    \draw [thick,->,gray] (10,0) -- (12,0);
    \node[align=center] at (1.45,-0.7) {$\sigma_0 = \{\}$};
    \draw [thick,decorate,decoration={brace,amplitude=6pt,raise=0pt,mirror}] (-0.25,-0.15) -- (3.15,-0.15);
    \node[align=center] at (4.95,-0.7) {$\sigma_1 = \{a\}$};
    \draw [thick,decorate,decoration={brace,amplitude=6pt,raise=0pt,mirror}] (3.25,-0.15) -- (6.65,-0.15);
    \node[align=center] at (8.35,-0.7) {$\sigma_2 = \{a,b\}$};
    \draw [thick,decorate,decoration={brace,amplitude=6pt,raise=0pt,mirror}] (6.75,-0.15) -- (9.95,-0.15);
    \node[align=center,gray] at (11.15,-0.7) {$\sigma_3 = \{b\}$};
    \draw [thick,decorate,decoration={brace,amplitude=6pt,raise=0pt,mirror},gray] (10.05,-0.15) -- (12.25,-0.15);
    \node[align=center] at (3.2,1.3) {$t=3.2$};
    \draw [thick,->] (3.2,1) -- (3.2,0.15);
    \node[align=center] at (6.7,1.3) {$t=6.7$};
    \draw [thick,->] (6.7,1) -- (6.7,0.15);
    \node[align=center] at (10,1.3) {$t=10$};
    \draw [thick,->] (10,1) -- (10,0.15);
    \node[align=center,gray] at (11.8,0.8) {\textbf{Violation}};
    \draw [line width=1.6pt,->,gray] (11.8,0.5) -- (10.3,0.15);
  \end{tikzpicture}
  \caption{A sample computation ($\sigma = \{\{\}, \{a\}, \{a,b\}, \{b\}\}$) resulting in the violation of $\phi$}
  \end{figure}
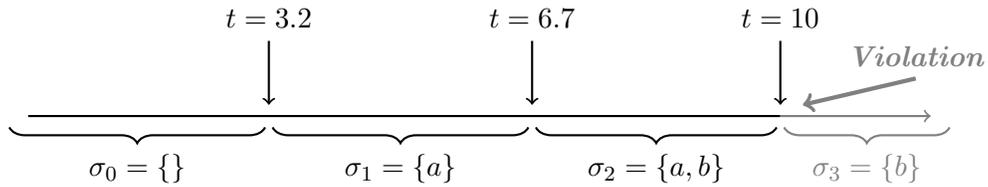

  \end{center}
\end{example}

\begin{definition}[Linear Temporal Logic (LTL)]
\label{def:LTL}
LTL is a popular property specification language, used to define correctness properties with the following syntax:
$$\phi ::= p\ |\ \lnot \phi\ |\ \phi \vee \phi\ |\ \bigcirc \phi\ |\ \phi \cup \phi$$
, where $p$ is an atomic proposition ($p \in AP$).
The semantics of LTL is based on infinite computations of the system, and is defined as follows.
Considering $\omega = \{\sigma_0, \sigma_1, \sigma_2, \dots\} \in \Sigma^\omega$, the satisfaction relation (depecited with $\models$) is defined as follows:
\begin{eqnarray*}
  \omega \models \phi &\ \textbf{iff}\ & \omega, \sigma_0 \models \phi \\
  \\
  \omega, \sigma_i \models p &\ \textbf{iff}\ & p \in \sigma_i \\
  \omega, \sigma_i \models \lnot \phi &\ \textbf{iff}\ & \omega \not\models \phi \\
  \omega, \sigma_i \models \phi \vee \psi &\ \textbf{iff}\ & (\omega, \sigma_i \models \phi) \vee (\omega, \sigma_i  \models \psi) \\
  \omega, \sigma_i \models \bigcirc \phi &\ \textbf{iff}\ & \sigma_{i+1} \models \phi \\
  \omega, \sigma_i \models \phi \cup \psi &\ \textbf{iff}\ & \exists k \geq i \;.\; (\omega,\sigma_k \models \psi) \wedge \forall i \leq j \leq k \;.\; (\omega,\sigma_j \models \phi)
\end{eqnarray*}

Temporal operators $\square$ and $\lozenge$ can also be defined as follows:

\begin{eqnarray*}
  \lozenge \phi &=& \top \cup \phi\\
  \square \phi &=& \lnot (\lozenge \lnot \phi)
\end{eqnarray*}
\end{definition}

The semantics of LTL is defined based on infinite computations and cannot be used in the context of runtime verification, as in that case, we are dealing with finite computations.
Therefore a finite variant of this language has been proposed, called LTL$_3$~\cite{ltl3}, the semantics of which is defined on finite computations.
The syntax of LTL$_3$ is exactly the same as LTL, but its semantics is different.

\begin{definition}[LTL$_3$ (3-valued LTL)]
\textit{LTL$_3$}, also known as \textit{3-valued LTL} is very similar to LTL, with slightly different semantics.
The evaluation of an LTL$_3$ property $\phi$ on a finite computation $\eta \in \Sigma^*$ is defined below. Note that the dot operator implies concatenation, and $\models$ used in this definition is already defined in Definition~\ref{def:LTL}.
$$[\eta \models \phi] \equiv \left\{
  \begin{tabular}{r l}
  $\top$ & iff $\forall \omega \in \Sigma^\omega:(\eta.\omega \models \phi)$\\
  $\bot$ & iff $\forall \omega \in \Sigma^\omega:(\eta.\omega \not\models \phi)$\\
  $?$ & otherwise
  \end{tabular}
\right.$$
In other words, a finite computation satisfies (violates) an LTL$_3$ property, iff for all extensions of that computation to an infinite one, the corresponding LTL property is satisfied (violated). Otherwise the property is evaluated to \textit{unknown} (denoted by $?$). Simply speaking, LTL is intended for offline verification of the system and results in a value in $ \{\top, \bot\}$ where $\top$ means \textit{true/satisfied} and $\bot$ means \textit{false/violated}. LTL$_3$, on the other hand, is intended to be evaluated at runtime on finite computations and may be evaluated to unknown as well.
\end{definition}

\begin{example}
In Example~\ref{eg:main}, $\phi$ can be defined as: $\phi = \lnot a \cup (a \cup (b \wedge c))$ as an LTL$_3$ formula.
\end{example}

Some properties like ``Drone B arrives at the destination and leaves it infinately often" cannot be checked using runtime verification.
These properties are also refered to as \textit{non-monitorable properties}.
Their counterparts, also refered to as \textit{monitorable properties}, are the types of properties that can benefit from our algorithm.

\begin{definition}[Good/Bad Prefix]
For a computation $\omega = \{\sigma_0, \sigma_1, \sigma_2, \dots\}$, and a given property $\phi$, the prefix $\omega' = \{\sigma_0, \sigma_1, \dots, \sigma_k\}$ is a good/bad prefix if it results in the satisfaction/violation of $\phi$.
\end{definition}

\noindent (Non-)Monitorable properties are formally defined as follows:
\begin{definition}[(Non-)Monitorable property]
An LTL property $\phi$ is monitorable/non-monitorable iff there exists/doesn't exist a computation $\omega$ containing a good or bad prefix of $\phi$.
\end{definition}

Monitorable properties are convertable to a specific DFA (deterministic finite automaton), which can be used in runtime verification.
How to convert an LTL property to its corresponding automaton is described in detail in \cite{ltl3,ltl3_2}.
In the following examples, we show how this automaton is useful.

\begin{definition}[Monitor Automaton]
For a monitorable LTL property $\phi$, the monitor automaton $M_{\phi} = \{Q,Q_0,R,L\}$ is a unique deterministic finite automaton (DFA), where:
\begin{itemize}
  \item $Q$ is a set of locations,
  \item $Q_0 \subseteq Q$ is a set of initial locations,
  \item $R \subseteq Q \times 2^\Sigma \times Q$ is a transition relation on the set of locations, and
  \item $L : Q \mapsto \{\top, \bot, ?\}$ is a labeling function that shows whether the location is accepting ($\top$), rejecting ($\bot$), or unknown ($?$).
\end{itemize}

For every finite computation $\eta$ that satisfies/violates $\phi$, the run of $M_{\phi}$ on $\eta$ terminates in a location labeled $\top$/$\bot$, and for every computation $\eta$ that neither satisfies nor violates $\phi$, the run terminates in a location that is labeled $?$ by the $L$ function.
For simplicity, the label of each transition is depicted by a predicate in propositional logic, which represents a subset of $\Sigma$, containing elements that can each enable the transition.
\end{definition}

\begin{example}
  The monitor automaton for the LTL property $\phi = \lnot a \cup (a \cup (b \wedge c))$ is depicted in Fig.~\ref{fig:automaton} ($\top$ and $\bot$ mean $\mathit{true}$ and $\mathit{false}$ respectively and are used to show satisfying/violating states):
  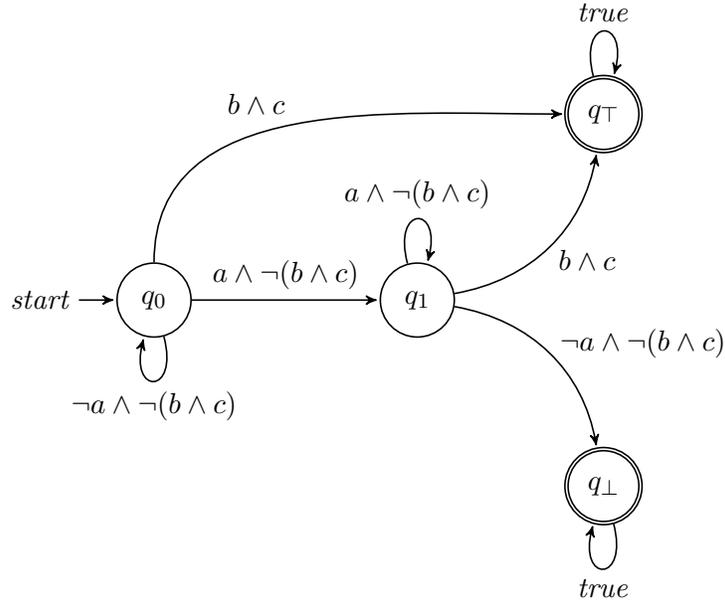
\begin{figure}[H]
    \begin{center}
    \begin{tikzpicture}[->,>=stealth',shorten >=1pt,auto,node distance=3.5cm,semithick]
      \tikzstyle{every state}=[fill=white,draw=black,text=black]
      \node[initial, state]   (S0) {$q_0$};
      \node[state]            (S1) [right of = S0] {$q_1$};
      \node[state, accepting] (ST) [above right of = S1] {$q_\top$};
      \node[state, accepting] (SF) [below right of = S1] {$q_\bot$};

      \path (S0)  edge  [loop below]              node                            {$\lnot a \wedge \lnot (b \wedge c)$}  (S0)
                  edge                            node                            {$a \wedge \lnot (b \wedge c)$}        (S1)
                  edge  [out=90, in=180]          node                            {$b \wedge c$}                         (ST)
            (S1)  edge  [loop above]              node                            {$a \wedge \lnot (b \wedge c)$}        (S1)
                  edge  [out=10, in=-100, below]  node  [xshift=5mm, yshift=1mm]  {$b \wedge c$}                         (ST)
                  edge  [out=-10, in=100]         node  [yshift=-2mm]             {$\lnot a \wedge \lnot (b \wedge c)$}  (SF)
            (ST)  edge  [loop above]              node                            {$\mathit{true}$}                      (ST)
            (SF)  edge  [loop below]              node                            {$\mathit{true}$}                      (SF);
    \end{tikzpicture}
    \end{center}
    \caption{The monitor automaton for the LTL property $\phi = \lnot a \cup (a \cup (b \wedge c))$}
    \label{fig:automaton}
  \end{figure}
\end{example}

A monitor automaton is deterministic, which means that for any location $q \in Q$ and any $\sigma \in \Sigma$, there exists $q' \in Q$ such that $(q,\sigma,q') \in R$ and $\forall {q'' \in Q} \; : \; (q,p,q'') \in R \Rightarrow q'' = q'$.
Simply speaking, at each point in time, one and exactly one transition is enabled (including self-loops).

\begin{example}
  Following Examples~\ref{eg:computations} and~\ref{eg:computationsWithAP}, we can show the changes in the locations of the corresponding monitor automation (Fig.~\ref{fig:automaton}) in Figs.~\ref{fig:computationWithLocations1} and \ref{fig:computationWithLocations2}, where the bottom braces show the current location of the automaton. Starting at state $q_0$, each time a transition is enabled based on the system's global state, the current location changes to the succeeding location accordingly.

  \begin{center}

  \begin{figure}[H]
  \begin{tikzpicture}[scale=1]
    \draw [thick] (0,0) -- (9,0);
    \draw [thick,->,gray] (9,0) -- (12,0);
    \node[align=center] at (0.9,-0.7) {$\{\}$};
    \draw [thick,decorate,decoration={brace,amplitude=6pt,raise=0pt,mirror}] (-0.25,-0.15) -- (2.05,-0.15);
    \node[align=center] at (3.65,-0.7) {$\{a\}$};
    \draw [thick,decorate,decoration={brace,amplitude=6pt,raise=0pt,mirror}] (2.15,-0.15) -- (5.15,-0.15);
    \node[align=center] at (7.1,-0.7) {$\{a,b\}$};
    \draw [thick,decorate,decoration={brace,amplitude=6pt,raise=0pt,mirror}] (5.25,-0.15) -- (8.95,-0.15);
    \node[align=center,gray] at (10.65,-0.7) {$\{a,b,c\}$};
    \draw [thick,decorate,decoration={brace,amplitude=6pt,raise=0pt,mirror},gray] (9.05,-0.15) -- (12.25,-0.15);
    \node[align=center] at (2.1,1.8) {$t=2.1$};
    \node[align=center,gray] at (2.1,1.3) {$a \wedge \lnot (b \wedge c)$};
    \draw [thick,->] (2.1,1) -- (2.1,0.15);
    \node[align=center] at (5.25,1.8) {$t=5.2$};
    \node[align=center,white!30!black] at (5.2,1.3) {\text{[self-loop]}};
    \draw [thick,->] (5.2,1) -- (5.2,0.15);
    \node[align=center] at (9,1.8) {$t=9$};
    \node[align=center,gray] at (9,1.3) {$b \wedge c$};
    \draw [thick,->] (9,1) -- (9,0.15);
    \node[align=center,gray] at (11,0.8) {\textbf{Satisfaction}};
    \draw [line width=1.6pt,->,gray] (11,0.5) -- (9.3,0.15);

    \node[align=center,gray] at (0.9,-1.7) {$q_0$};
    \draw [thick,decorate,decoration={brace,amplitude=6pt,raise=0pt,mirror},gray] (-0.25,-1.15) -- (2.05,-1.15);
    \node[align=center,gray] at (5.55,-1.7) {$q_1$};
    \draw [thick,decorate,decoration={brace,amplitude=6pt,raise=0pt,mirror},gray] (2.15,-1.15) -- (8.95,-1.15);
    \node[align=center,gray] at (10.65,-1.7) {$q_\top$};
    \draw [thick,decorate,decoration={brace,amplitude=6pt,raise=0pt,mirror},gray] (9.05,-1.15) -- (12.25,-1.15);
  \end{tikzpicture}
  \caption{A sample execution scenario resulting in the satisfaction of $\phi$}
  \label{fig:computationWithLocations1}
  \end{figure}
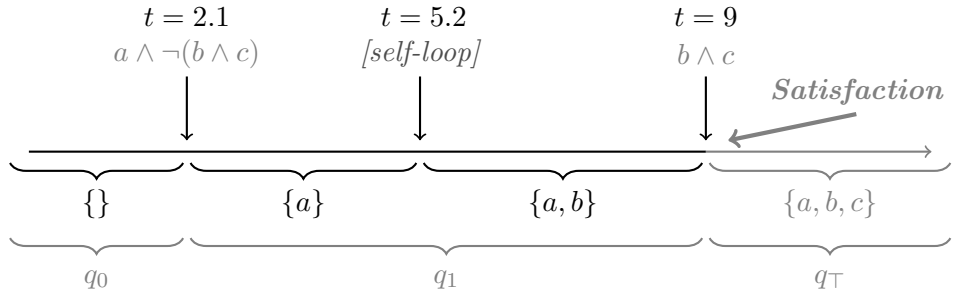

  \begin{figure}[H]
  \begin{tikzpicture}[scale=1]
    \draw [thick] (0,0) -- (10,0);
    \draw [thick,->,gray] (10,0) -- (12,0);
    \node[align=center] at (1.45,-0.7) {$\{\}$};
    \draw [thick,decorate,decoration={brace,amplitude=6pt,raise=0pt,mirror}] (-0.25,-0.15) -- (3.15,-0.15);
    \node[align=center] at (4.95,-0.7) {$\{a\}$};
    \draw [thick,decorate,decoration={brace,amplitude=6pt,raise=0pt,mirror}] (3.25,-0.15) -- (6.65,-0.15);
    \node[align=center] at (8.35,-0.7) {$\{a,b\}$};
    \draw [thick,decorate,decoration={brace,amplitude=6pt,raise=0pt,mirror}] (6.75,-0.15) -- (9.95,-0.15);
    \node[align=center,gray] at (11.15,-0.7) {$\{b\}$};
    \draw [thick,decorate,decoration={brace,amplitude=6pt,raise=0pt,mirror},gray] (10.05,-0.15) -- (12.25,-0.15);
    \node[align=center] at (3.2,1.8) {$t=3.2$};
    \node[align=center,gray] at (3.2,1.3) {$a \wedge \lnot (b \wedge c)$};
    \draw [thick,->] (3.2,1) -- (3.2,0.15);
    \node[align=center] at (6.7,1.8) {$t=6.7$};
    \node[align=center,white!30!black] at (6.7,1.3) {\text{[self-loop]}};
    \draw [thick,->] (6.7,1) -- (6.7,0.15);
    \node[align=center] at (10,1.8) {$t=10$};
    \node[align=center,gray] at (10,1.3) {$\lnot a \wedge \lnot (b \wedge c)$};
    \draw [thick,->] (10,1) -- (10,0.15);
    \node[align=center,gray] at (11.8,0.8) {\textbf{Violation}};
    \draw [line width=1.6pt,->,gray] (11.8,0.5) -- (10.3,0.15);

    \node[align=center,gray] at (1.45,-1.7) {$q_0$};
    \draw [thick,decorate,decoration={brace,amplitude=6pt,raise=0pt,mirror},gray] (-0.25,-1.15) -- (3.15,-1.15);
    \node[align=center,gray] at (6.6,-1.7) {$q_1$};
    \draw [thick,decorate,decoration={brace,amplitude=6pt,raise=0pt,mirror},gray] (3.25,-1.15) -- (9.95,-1.15);
    \node[align=center,gray] at (11.15,-1.7) {$q_\bot$};
    \draw [thick,decorate,decoration={brace,amplitude=6pt,raise=0pt,mirror},gray] (10.05,-1.15) -- (12.25,-1.15);
  \end{tikzpicture}
  \caption{A sample execution scenario resulting in the violation of $\phi$}
    \label{fig:computationWithLocations2}
  \end{figure}
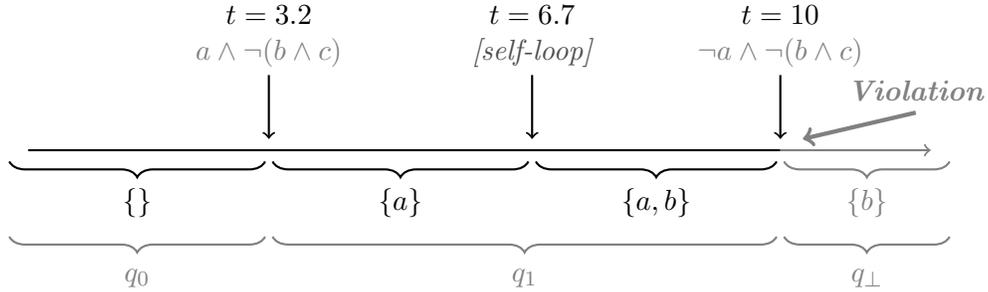

  \end{center}


\end{example}

Essentially, we need to present a decentralized algorithm that detects the location changes in the monitor automaton by monitoring changes in the system's global state.
For simplicity, we modify the automaton by first removing self-loops and then replacing the transition predicates with their corresponding Disjunctive Normal Form (DNF).

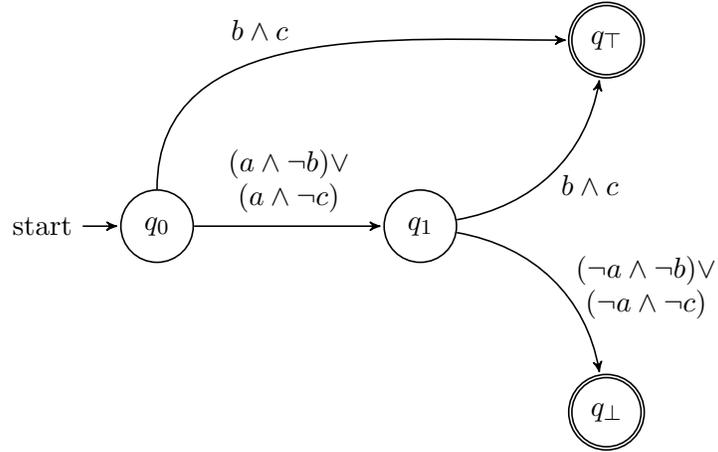
\begin{figure}[H]
\begin{center}
\begin{tikzpicture}[->,>=stealth',shorten >=1pt,auto,node distance=3.5cm,semithick]
  \tikzstyle{every state}=[fill=white,draw=black,text=black]
  \node[initial, state]   (S0) {$q_0$};
  \node[state]            (S1) [right of = S0] {$q_1$};
  \node[state, accepting] (ST) [above right of = S1] {$q_\top$};
  \node[state, accepting] (SF) [below right of = S1] {$q_\bot$};
  \path (S0) edge node {\begin{tabular}{c}$(a \wedge \lnot b) \vee$ \\ $(a \wedge \lnot c)$\end{tabular}} (S1)
             edge [out=90, in=180] node {$b \wedge c$} (ST)
        (S1) edge [out=10, in=-100, below] node [xshift=5mm, yshift=1mm] {$b \wedge c$} (ST)
             edge [out=-10, in=100] node [yshift=-7mm] {\begin{tabular}{c}$(\lnot a \wedge \lnot b) \vee$ \\ $(\lnot a \wedge \lnot c)$\end{tabular}} (SF);
\end{tikzpicture}
\end{center}
\caption{The modified monitor automaton for the LTL property $\phi = \lnot a \cup (a \cup (b \wedge c))$}
\end{figure}

We further modify the automaton by breaking transitions into multiple transitions, each containing one conjunct from the DNF form of the transition predicate. We also label each transition by $Tr_i$.

\begin{figure}[H]
\begin{center}
\begin{tikzpicture}[->,>=stealth',shorten >=1pt,auto,node distance=3.5cm,semithick]
  \tikzstyle{every state}=[fill=white,draw=black,text=black]
  \node[initial, state]   (S0) {$q_0$};
  \node[state]            (S1) [right of = S0] {$q_1$};
  \node[state, accepting] (ST) [above right of = S1] {$q_\top$};
  \node[state, accepting] (SF) [below right of = S1] {$q_\bot$};
  \path (S0) edge [out=30, in=150] node                         {$Tr_0 = a \wedge \lnot b$} (S1)
             edge [out=-30, in =-150, below] node               {$Tr_1 = a \wedge \lnot c$} (S1)
             edge [out=90, in=180] node                         {$Tr_2 = b \wedge c$} (ST)
        (S1) edge [out=10, in=-100, below] node [xshift=9mm]    {$Tr_3 = b \wedge c$} (ST)
             edge [out=-10, in=100] node [yshift=-1mm]          {$Tr_4 = \lnot a \wedge \lnot b$} (SF)
             edge [out=-90, in=180, below] node [xshift=-11mm]  {$Tr_5 = \lnot a \wedge \lnot c$} (SF);
\end{tikzpicture}
\end{center}
\caption{The final monitor automaton for the LTL property $\phi = \lnot a \cup (a \cup (b \wedge c))$}
\label{eg:finalAutomaton}
\end{figure}
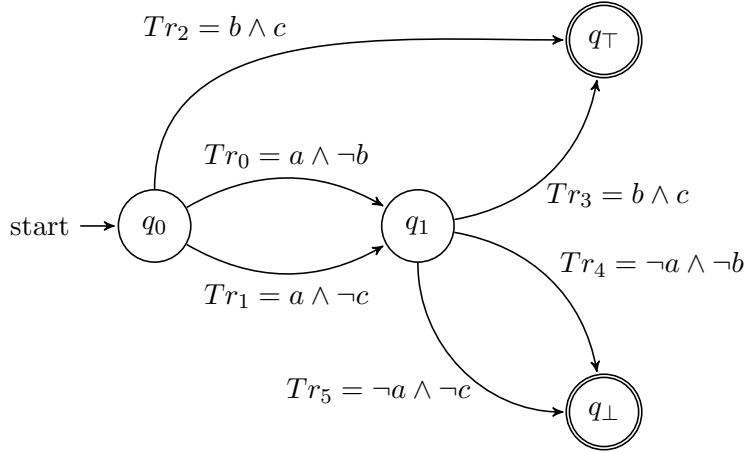

The automaton depicted in Fig.~\ref{eg:finalAutomaton} is what we would be using throughout our monitoring algorithm; at each point in time, the whole system is at one of the automaton locations $\{q_0, q_1, q_\top, q_\bot\}$.
Note that in a scenario where our monitor automaton location is $q_0$, and the system's global state is $\{a\}$, then both $Tr_0$ and $Tr_1$ would be enabled.
A similar scenario could be shown for $Tr_4$ and $Tr_5$ where the monitor automaton location is $q_1$, and the system's global state is $\{\}$, however in all these cases, because of the automaton's deterministic nature, all the transitions arrive at the same succeeding location, and hence, it doesn't matter whether for example, $Tr_0$ gets enabled before $Tr_1$, or vice versa.

\section{Problem Statement}
\label{sect:problem}

We take as input: 
\begin{enumerate}
	\item a distributed system $S$, consisting of $n$ processes, $\{p_1, p_2, p_3, \dots, p_n\}$ sharing a global clock and connecting via asynchronous message passing, and
	\item an LTL property $\phi$.
\end{enumerate}
We propose an algorithm for a local monitor $m_i$ for each process $p_i$ to monitor $\phi$ on $S$, such that its violation or satisfaction is detected by at least one process.
The same exact algorithm would be run as each process's monitor.
The only differentiating factors between $m_i$'s is the set of local variables that they can read, which belongs to the process $p_i$.
We assume that there is a message channel between each pair of processes, and if a message is sent, it will be received by its intended receiver in finite time and not get dropped, corrupted or tampered with.

\section{Decentralized algorithm}
\label{sect:algorithm}

In this section, we will present our decentralized algorithm for runtime verification in detail.
The algorithm takes an LTL property as input. The property's monitor automaton is initially constructed offline, having states $\{q_0, q_1, q_2, \dots\}$, and transitions $\{Tr_1, Tr_2, Tr_3, \dots\}$, each having an expression as its predicate, which is a propositional conjunction of atomic propositions. Each process keeps a copy of this automaton.

\noindent \textbf{Note:} From this point in the paper, when will use  processes and their monitors interchangeably.

The general goal of the monitoring algorithm is to trace the locations of the monitor automaton based on the changes in the global state of the system, and detect reaching $\top$ or $\bot$ locations.
For instance, in our running example, for the computation $\eta=\{\{\}, \{b\}, \{a,b\}, \{a\},\{\}\}$, the corresponding automaton locations of the system are $q_0$ for the first two letters (global states), $q_1$ for the next ($\{a, b\}$), and $q_\bot$ for the last one ($\{\}$).

At each step, we want to detect the change from one automaton location to the other. In other words, among all the outgoing transitions from the current automaton location, we want to detect the one that gets enabled the earliest.

\subsection{Detecting enabling time of transitions}

Each process stores the current automaton location ($q_{current}$, or $q_c$ for short).
Initially, this location is the initial automaton location ($q_0$).
Each time a process detects the location change, it updates its own $q_c$ and moves on to the next step.
The last location change time is also stored by each process in its $t_{last\_location\_change}$ (or $t_{llc}$ for short) value, which is initially zero. Note that location changes may be detected with a slight delay, for example, if at $t = 7$ the system arrives at a new automaton location, this fact may remain undetected for 2 seconds until at $t = 9$ one of the processes realizes that at $t = 7$ a transition was enabled (and it was the earliest among the others) resulting in an automaton location change. That process then lets the others know about this location change.

In order to detect a transitions enabling time, the truthfulness of its predicate must be monitored. Each process may have one or more \textbf{literals} in a predicate. If, for example, a transition's predicate is $(\lnot a_1 \wedge a_2 \wedge \lnot b)$, and $a_1$ and $a_2$ are only accessible by process $p_1$, and $b$ is only accessible by process $p_2$, then $\lnot a_1$ and $a_2$ are the literals of process $p_1$, and $\lnot b$ is the literal of process $p_2$. Given that assumption, if $a_1 = true$ and $a_2 = false$, we say $p_1$'s literals are satisfied, and also if $b = false$, we say $p_2$'s literals are satisfied. Therefore, a transition becomes enabled when for each process associated with the transition's predicate, all the process's literals get satisfied.

At each step, for each outgoing transition of $q_c$, all processes associated with the transition's predicate are responsible for detecting if and when it becomes enabled. Each of these processes records a history of when its literals were satisfied from $t_{llc}$ until now. This history is stored as a set of positive intervals and we call it the $\mathit{satisfaction\_range}$ of that process for that transition, or $sr$ for short. For example,  $sr = [3, 4] \cup [5, 6]$ indicates that at all times between $t = 3$ and $t = 4$, and also at all times between $t = 5$ and $t = 6$, all the process's literals were satisfied for the transition. Each process keeps its $sr$ up to date for all transitions whose predicates it is associated with.

As stated previously, in order for a transition to become enabled, the literals for all the processes associated with its predicate must be satisfied, and hence, a process can not make this decision on its own. Therefore, each process must keep the values of $sr$ for all the other processes associated with the transition's predicate. To this end, for each outgoing transition of $q_c$, if a process is associated with the transition's predicate, it stores the set $\{sr_1, sr_2, sr_3, \dots, sr_k\}$, containing the values of $sr$ for all the $k$ processes associated with the transition's predicate, including itself. However, we know that a process doesn't have access to the value of other process's $sr$; so, processes must somehow send their $sr$ values to each other. If a process realizes that $\mathit{global\_satisfaction\_range} = sr_1 \cap sr_2 \cap sr_3 \cap \dots \cap sr_k$ is not empty, then it can conclude that the transition was enabled at the smallest time in $\mathit{global\_satisfaction\_range}$, or $gsr$ for short. In other words, if $gsr \neq \O$ then $min(gsr)$ is the transition's enabling time.

In order for the values of $sr$ to be shared among the processes associated with all transitions' predicates, each transition has a \textbf{coordinator} process. At the beginning of each step, for each outgoing transition of $q_c$ a process is chosen to be its coordinator. The values of $sr$ for all processes are also set to zero at the beginning of a step. Whenever the coordinator realizes that one or more of its own literals are not satisfied in the transition at some points in time, it can conclude that the transition was definitely not enabled in those times, due to the transition's predicate being a conjunction of literals. However, when all of the coordinator's literals are satisfied, other processes must be consulted in the decision making, so the coordinator sends a \textbf{Delegate} message to another process associated with the transition's predicate. This message is structured as follows:
$$\langle Delegate, t_{llc}, Tr, \{sr_1, sr_2, sr_3, \dots, sr_k\}\rangle$$
$Tr$ in this message is the transition whose satisfaction is being determined, and the value of each $sr_i$ is the last value of $sr$ for $p_i$ that the coordinator knows of. The value of $t_{llc}$ is included in every message of any kind. It acts as a timestamp to allow us to distinguish between messages belonging to different steps; if a process receives a message with a smaller $t_{llc}$ than its own, it concludes that the message belongs to a previous step and is of no value, and hence, it is dropped by the process. If the message's $t_{llc}$ is greater than of the process, it means that the process is lagging behind other processes, and hence, it resets itself, updates its $t_{llc}$, processes the message, and from then after, waits for messages with the new value of $t_{llc}$.

Upon receiving a Delegate message, the receiving process becomes the new coordinator of the transition and the previous coordinator that sent the message no longer considers itself the coordinator of that transition. After receiving the message, the new coordinator updates its values of $sr$ belonging to other processes, to the corresponding values provided in the message. It then checks if an enabling time can be detected for the transition, just like the previous coordinator used to do. Hence, the coordination role keeps getting delegated between processes. The processes keep sharing their values of $sr$ with each other until one of them detects the transition's enabling time (if it ever becomes enabled). After detecting a transition's enabling time, the transition's coordinator will not send a Delegate message to the other processes in that transition, until the end of that step. Note that if a process is associated with the predicate of multiple outgoing transitions of $q_c$, it may become the coordinator of multiple transitions; so a process may be the coordinator of more than one transition at each point in time.

If a process detects that, for example, transition $Tr_i$ was enabled at a certain time, that does not mean that $Tr_i$ was the earliest transition to become enabled; another transition $Tr_j$ may exist with a smaller enabling time than $Tr_i$. Therefore, processes must check all outgoing transitions and determine which of them gets enabled at the earliest time. To this end, each process keeps a list called $transitions\_checked$, or $TrC$ for short, containing outgoing transitions it has checked so far. At the beginning of each step, this list is empty for all processes, and each time a process detects the enabling time of a transition, it adds that transition to its $TrC$ list. Each process also keeps track of the earliest transition that it has detected so far; the first time that a process detects a transition becoming enabled, it keeps that transition in its $Tr_{earliest}$ value ($Tr_e$ for short), and it keeps the transition's  enabling time in its $t_{Tr_e}$ value. From then after, whenever the process detects the enabling time of another transition, it adds that transition to $TrC$ and compares its enabling time to $t_{Tr_e}$; if the enabling time was smaller, the process updates its values of $Tr_e$ and $t_{Tr_e}$.

A transition may become enabled a long time after others (or not become enabled at all). For these transitions, it suffices for processes to detect that they won't be enabled earlier than their $Tr_e$; if the lower bound of a transition's enabling time is greater than a process's $t_{Tr_e}$, that process can be certain that the transition will never replace its $Tr_e$. To have a  lower bound for a transition's enabling time, we introduce a transition's \textit{last update time}, denoted by $t_{last\_updated\_by\_p_i}$, or $t_{lu_i}$ for short.  Each process keeps the last time it has received a Delegate message from each process $p_i$ as its $t_{lu_i}$ value. The minimum of all $t_{lu_i}$s is the lower bound of the transitions enabling time, since all the processes associated with the transition's predicate have updated the $sr$ value before that time, and if the transition was enabled before then, it would have been noticed. The set of all of a process's $t_{lu}$ values is sent with each Delegate message; so the structure of a Delegate message is as follows:
$$\langle Delegate, t_{llc}, Tr, \{sr_1, sr_2, sr_3, \dots, sr_k\}, \{t_{lu_1}, t_{lu_2}, t_{lu_3}, \dots, t_{lu_k}\} \rangle$$
Upon receiving a Delegate message, the next coordinator updates its own $t_{lu}$ values to the ones in the message. If after updating all its $sr$ and $t_{lu}$ values, it still can't detect the $Tr$'s enabling time, it checks if $min(\{t_{lu_1}, t_{lu_2}, t_{lu_3}, \dots, t_{lu_k}\}) < t_{Tr_e}$; if so, it means that $Tr$ was not enabled before $Tr_e$, therefore the process adds $Tr$ to its $TrC$ and no longer sends a Delegate message for $Tr$. Essentially, $TrC$ is the list of all transitions that have been checked and were not enabled earlier than $t_{Tr_e}$.

The values of $t_{lu}$ can also be used to determine the next coordinator process. More accurately, the process $p_j$ with the smallest value for $t_{lu_j}$ is selected, as it has sent its Delegate message before the others and is the least up-to-date one.  $t_{lu}$ values can also be used in a way that we no longer need to store an $sr$ value for each process. We can keep a $global\_potential\_satisfaction\_range$ value, or $gpsr$ for short, containing times when the transition \textbf{could have been} enabled. This value has the following relationship with $sr$ and $t_{lu}$ values:
$$gpsr = (sr_1 \cup (t_{lu_1}, \infty)) \cap (sr_2 \cup (t_{lu_2}, \infty)) \cap (sr_3 \cup (t_{lu_3}, \infty)) \cap \dots \cap (sr_k \cup (t_{lu_k}, \infty))$$
So $gpsr$ is similar to $sr$, except that we make the optimistic assumption that the literals of each other process are satisfied from that process's $t_{lu}$ (the last time we know about that processes literals) onward.
At the start of each step the value of $gpsr$ is $[0, \infty)$ in each process for each transition. The coordinator keeps the value of $gpsr$ up-to-date by removing the times when one or more of their literals were not satisfied from  $gpsr$. At each point in time, if $gpsr$ is not empty and $min(gpsr)$ is smaller than all the $t_{lu}$ values, then $t = min(gpsr)$ is the enabling time of the transition; this can be proven as follows:
$$gsr = \bigcap_{i=1}^k (sr_i) = \bigcap_{i=1}^k ((sr_1 \cup (t_{lu_1}, \infty)) \cap [0, t_{lu_1}]) = gpsr \cap \bigcap_{i=1}^k [0, t_{lu_i}]$$
$$\Rightarrow gsr = gpsr \cap [0, min(\{ t_{lu_1}, t_{lu_2}, t_{lu_3}, \dots, t_{lu_k} \})]$$
The enabling time for the transition is the minimum of $gsr$ values, which based on the equation above, is also the minimum of $gpsr$ values, given that this minimum is smaller than all values of $t_{lu}$. If $t = min(gpsr)$ was greater than one or more $t_{lu}$ values, it means that the transition may have become enabled but we lack data from the corresponding processes of those $t_{lu}$ values. Therefore, the coordinator sends a Delegate message to the next coordinator which has the smallest $t_{lu}$, which is among those ambiguous processes.

Having the values $gpsr$ and $t_{lu}$ of all the processes associated with the transition, we do not need to send $sr$ values with Delegate messages, so the final form of Delegate messages, which is used in the algorithm is structured as below:
$$\langle Delegate, t_{llc}, Tr, gpsr, \{t_{lu_1}, t_{lu_2}, t_{lu_3}, \dots, t_{lu_k}\}\rangle$$

\begin{example}
	Consider a distributed system including processes $\{p_1, p_2, p_3, p_4\}$ with $AP_1=\{a\},\ AP_2=\{b\},\ AP_3=\{c\},$ and $AP_4=\{d\}$. Assume that we want to detect the enabling time of a transition $Tr$ with predicate $a \wedge b \wedge c \wedge d$.
	
	Fig.~\ref{fig:timeline} shows a hypothetical timeline where the horizontal highlighted boxes indicate the intervals when each atomic proposition gets $\mathit{true}$ and vertical striped areas represent time ranges where $a \wedge b \wedge c \wedge d = true$: $[16, 18)$ and $[19, 20)$. We only care about the first time point in these ranges, as it is the enabling time of the transition.
	
	\begin{figure}[H]
		\begin{center}
			\includegraphics[width=\linewidth]{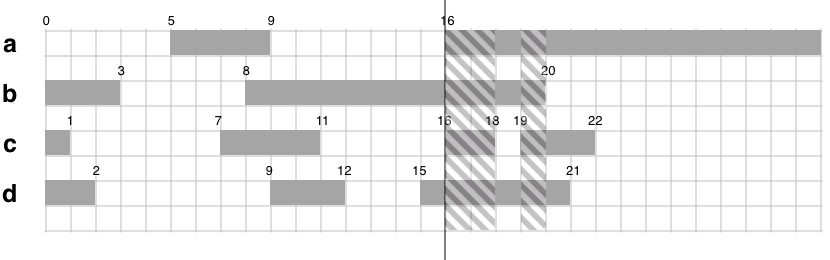}
		\end{center}
		\caption{An example of a  timeline showing the intervals that  each atomic proposition gets $\mathit{true}$}
		\label{fig:timeline}
	\end{figure}
	
	\noindent A sample execution of the transition monitoring part of our algorithm is depicted below:
	\begin{itemize}
		\item Initially, the values of $a$, $b$, $c$ and $d$ are $\mathit{false}$, and $p_1$ is the coordinator of the transition.
		\item At $t = 5$, $a$ becomes $\mathit{true}$ in process $p_1$. As a result, $p_1$ sends a Delegate message to $p_2$ with $gpsr = \{5\}$ and $t_{lu_1} = 5$.
		\item At $t = 6$, $p_2$ receives the Delegate message and updates its $gpsr$ to $\{\}$ (since $b$ is $\mathit{false}$ at $t = 5$). It also updates its $t_{lu_1}$ to $5$, and its $t_{lu_2}$ to $6$.
		\item At $t = 8$, $b$ becomes $\mathit{true}$ and $p_2$ sends a Delegate message to $p_3$ with $gpsr = \{8\}$, $t_{lu_1} = 5$, and $t_{lu_2} = 8$.
		\item At $t = 9$, $p_3$ receives the message and sets its $gpsr$ to $[8, 9]$, its $t_{lu_1}$ to $5$, its $t_{lu_2}$ to $8$, and its $t_{lu_3}$ to $9$. It then sends a Delegate message containing these values to $p_4$.
		\item At $t = 10$, $p_4$ receives the message, sets its $gpsr$ to $[9, 10]$, its $t_{lu_1}$ to $5$, its $t_{lu_2}$ to $8$, its $t_{lu_3}$ to $9$, and its $t_{lu_4}$ to $10$. It then sends a Delegate message containing these values to $p_1$.
		\item At $t = 11$, $p_1$ receives the message, sets its $gpsr$ to $\{\}$, its $t_{lu_1}$ to $11$, its $t_{lu_2}$ to $8$, its $t_{lu_3}$ to $9$, and its $t_{lu_4}$ to $10$.
		\item At $t = 16$, $a$ becomes $\mathit{true}$, so $p_1$ updates its $t_{lu_1}$ to $16$, and its $gpsr$ to $\{16\}$. It then sends a Delegate message to $p_2$.
		\item At $t = 16$, $p_2$ receives the message, sets its $gpsr$ to $[16, 19]$, its $t_{lu_1}$ to $16$, its $t_{lu_2}$ to $19$, its $t_{lu_3}$ to $9$, and its $t_{lu_4}$ to $10$. It then sends a Delegate message to $p_3$.
		\item At $t = 20$, $p_3$ receives the message, sets its $gpsr$ to $[16, 18] \cup [19, 20]$, its $t_{lu_1}$ to $16$, its $t_{lu_2}$ to $19$, its $t_{lu_3}$ to $20$, and its $t_{lu_4}$ to $10$. It then sends a Delegate message to $p_4$.
		\item At $t = 22$, $p_4$ receives the message, sets its $gpsr$ to $[16, 18] \cup [19, 21]$, its $t_{lu_1}$ to $16$, its $t_{lu_2}$ to $19$, its $t_{lu_3}$ to $20$, and its $t_{lu_4}$ to $22$. It then announces $t = 16$ as the transition's enabling time.
	\end{itemize}
\qedsymbol
\end{example}

\subsection{Issues Regarding Completeness}

As stated previously, processes detect the enabling times of transitions by sending Delegate messages to each other. They then add these transitions, along with other transitions that have been proven not to become enabled before others, to their $TrC$ list. Whenever a process arrives at a state where its $TrC$ contains all outgoing transitions of $q_c$, then its $Tr_e$ is the earliest transition and that transition's target location is the next automaton location. The process informs the other processes of this location change and the next step starts. One problem is that some transitions may never get added to a process's $TrC$ using the previously described methods. These transitions are present in one of the following scenarios:
\begin{enumerate}
	\item Consider the outgoing transitions of $q_c$ to be $Tr_1$ with predicate $a \wedge b$, and $Tr_2$ with predicate $c \wedge d$, where $a$, $b$, $c$, and $d$ are literals belonging to four different processes. The two processes that own $a$ and $b$ only communicate with each other about $Tr_1$, since they are not associated with $Tr_2$'s predicate, therefore by only sending and receiving Delegate messages, they will never be able to add $Tr_2$ to their $TrC$ lists. The same is true for the two processes that own $c$ and $d$, and transition $Tr_1$.
	\item Consider the outgoing transitions of $q_c$ to be $Tr_1$ with predicate $a \wedge b$, and $Tr_2$ with predicate $a \wedge c$, where $a$, $b$ and $c$ are literals belonging to $p_1$, $p_2$, and $p_3$, respectively.  Assume $t = 5$ is detected as $Tr_1$'s enabling time by $p_1$. If $p_3$ is currently the coordinator of $Tr_2$, and if $c$ never gets satisfied ($c$ is always equal to $\mathit{false}$), $p_3$ never sends a Delegate message to $p_1$, since its $gpsr = \O$. This is while $p_1$ only needs to know that $Tr_2$ never gets enabled before $t = 5$ in order to add it to its $TrC$ and come to the conclusion that $Tr_1$ is the earliest transition.
\end{enumerate}

In order to fix the aforementioned problem, we introduce a new kind of message called \textbf{Aggregate}. Whenever a process $p_i$'s $TrC$ list gets updated, it sends an Aggregate message, containing its $TrC$, $Tr_e$ and $t_{Tr_e}$, to processes that need information about transitions that have been checked. These receiving processes are defined in the next paragraph. Intuitively speaking, the receiving processes pass the list of checked transitions to other processes via Aggregate messages, until all the processes that need to know about the transitions checked so far are notified.

Having $TrC'$ be the list of $q_c$'s outgoing transitions that are not members of $p_i$'s $TrC$, $p_i$ sends this message to processes with at least one of the following conditions:
\begin{enumerate}
	\item Are associated with a transition in $TrC'$, the predicate of which $p_i$ is not associated with,
	\item Are associated with a transition in $TrC'$, the predicate of which $p_i$ is associated with, and their $t_{lu}$ in $p_i$ is smaller than $p_i$'s $t_{Tr_e}$.
\end{enumerate}

\noindent Aggregate message have the following structure:
$$\langle Aggregate, t_{llc}, TrC, Tr_e, t_{Tr_e} \rangle$$

When a process receives an Aggregate message, it compares the message's $t_{Tr_e}$ to its own  $t_{Tr_e}$, and if the message's $t_{Tr_e}$ is smaller than its own $t_{Tr_e}$, then it updates its $Tr_e$ and $t_{Tr_e}$ to the ones of the received Aggregate message. It also adds the transitions in the received message's $TrC$ to its own $TrC$ list. If, as a result, its $TrC$ list changes, and its new $TrC$ still does not contain all outgoing transitions of $q_c$, it sends an Aggregate message to other processes that have one of the previously mentioned conditions. Otherwise, it announces an automaton location change.

\subsection{Automaton Location Change}
When a process realizes a location change in an automaton, it notifies the coordinators of the outgoing transitions of the next automaton location. The initial coordinator of each transition is  the process with the smallest index; for example, for a transition with a predicate involving $p_1$, $p_2$ and $p_3$ in it, process $p_1$ is chosen as the coordinator. This criteria can be changed arbitrarily, but the important point is that all processes should have the same criteria for this choice, so that if more than one process notices this change at the same time, different coordinators are not notified for the same transition. Upon receiving the notification, the next chosen coordinator processes reset all their values and then next step of the algorithm starts.

A concurrency issue may occur, where more than one process realizes the location change, and they all send messages to the next coordinators. In this case, a coordinator may receive the notification message, delegate its role to another process, and receive another notification again from another process. This can trivially be fixed if the receiving coordinator process ignores the second notification message with the same $t_{llc}$ value as the previous one.\\

\subsection{Algorithm Overview}

A general overview of the algorithm is provided as pseudo-code in Algorithm~1 and Algorithm~2 (where $i$ is the index of the process that the algorithm is running on).

Algorithm~1 shows the routines that are run whenever a message is received by another monitor, and whenever $p_i$'s local state changes. Lines 2-3 drop the message if it belongs to a previous step. In lines 5-15, the monitor is reset, if it discovers from the value of $t_{llc}$ that a new step has begun. In lines 17-21, the process's $gpsr$ and $t_{lu}$ values are updated in the case of receiving a Delegate message, and in lines 23-28, $TrC$, $Tr_e$ and $t_{Tr_e}$ are updated in the case of receiving an Aggregate message. In line 30, the \texttt{UpdateMonitorState()} function is called, which is responsible for updating the monitors' local state and sending messages if necessary. This function is all that needs to happen when $p_i$'s local state changes (line 34).

\begin{algorithm}[H]
\caption{Message Reception and State Change Handlers for Each Process $p_i$}
\begin{algorithmic}[1]
\Procedure{OnReceiveMessage}{m}
  \If{$m.t_{llc} < t_{llc}$}
    \State // drop message (do nothing)
  \Else
    \If{$m.t_{llc} > t_{llc}$}
      \State // reset the process's monitoring state for the new location
      \State $t_{llc} \gets m.t_{llc}$
      \State $TrC \gets \{\}$
      \State $Tr_e \gets \bot$
      \State $t_{Tr_e} \gets 0$
      \ForAll{$Tr \in q_c.outgoing\_transitions$ \textbf{s.t.} $p_i \in Tr$}
        \State $isCoordinatorOf[Tr] \gets false$
        \State $Transitions.add(Tr, gpsr = 0, t_{lu} = [\ t_{llc}$ for all processes $])$
      \EndFor
    \EndIf
    \State $TrC\_has\_changed \gets false$
    \If{$m.type = Delegate$}
      \State $Tr \gets m.Tr$
      \State $Tr.gpsr \gets m.gpsr$
      \State $Tr.lu \gets m.lu$
      \State $isCoordinatorOf[Tr] \gets true$
    \Else
      \State $TrC \gets TrC \cup m.TrC$
      \State $TrC\_has\_changed \gets true$
      \If{$m.t_{Tr_e} < t_{Tr_e}$}
        \State $Tr_e \gets m.Tr$
        \State $t_{Tr_e} \gets m.t_{Tr_e}$
      \EndIf
    \EndIf
    \State \Call{$UpdateMonitorState()$}{}
  \EndIf
\EndProcedure
\Procedure{onLocalStateChange()}{}
  \State \Call{$UpdateMonitorState()$}{}
\EndProcedure
\end{algorithmic}
\end{algorithm}

Algorithm~2 shows the \texttt{UpdateMonitorState()} function. In lines 3-4, the values of the process's $t_{lu}$ and $gpsr$ are updated for each outgoing transition, along with adding the transitions, the enabling times of which have been detected, to $TrC$ in line 7 (the \texttt{RemoveUnsatisfactoryIntervals} helper function removes times were at least one if $p_i$'s literals are not satisfied, from the given $gpsr$). Lines 9-12 update the earliest outgoing transition ($Tr_e$) detected so far. Lines 17-20 add the transitions that can't be enabled before $Tr_e$ to $TrC$. Lines 21-26 send Delegate messages for the transitions which $p_i$ is the coordinator of, and also may be enabled up until now.

If $TrC$ is updated to contain all transitions, the algorithm will announce the monitor location change at line 30. Lines 33-48 send Aggregate messages (in the case of $TrC$ being changed) to processes associated with transitions that don't exist in $TrC$ and have the first (lines 36-38) or the second (lines 40-42) conditions described previously.

\begin{algorithm}[H]
\caption{The UpdateMonitorState function}
\begin{algorithmic}[1]
\Procedure{UpdateMonitorState()}{}
  \ForAll{$Tr \in q_c.outgoing\_transitions$ \textbf{s.t.} $p_i \in Tr$}
    \State $RemoveUnsatisfactoryIntervals(Tr.gpsr)$
    \State $Tr.lu_i = now$
    \If{$min(Tr.gpsr) < now$}
      \If{$\forall j: (Tr.lu_j > min(Tr.gpsr))$}
        \State $Trc.add(Tr)$
        \State $TrC\_has\_changed \gets false$
        \If{$min(Tr.gpsr) < t_{Tr_e}$}
          \State $Tr_e \gets Tr$
          \State $t_{Tr_e} \gets min(Tr.gpsr)$
        \EndIf
      \EndIf
    \EndIf
  \EndFor
  \ForAll{$Tr \in q_c.outgoing\_transitions$ \textbf{s.t.} $p_i \in Tr$}
    \If{$Tr.gpsr \cap [0, t_{Tr_e}) = \O \vee \forall j: (Tr.lu_j > t_{Tr_e})$}
      \State $Trc.add(Tr)$
      \State $TrC\_has\_changed \gets false$
    \EndIf
    \If{$Tr \not\in TrC \wedge min(Tr.gpsr) < now$}
        \If{$isCoordinatorOf[Tr] = true$}
          \State \textbf{send} $\langle Delegate, t_{llc}, r.gpsr, Tr.lu \rangle$ to $p_j$ with smallest $Tr.lu_j$
          \State{$isCoordinatorOf[Tr] \gets false$}
        \EndIf
    \EndIf
  \EndFor
  \If{$TrC\_has\_changed = true$}
    \If{$\forall Tr \in q_c.outgoing\_transitions: Tr \in TrC$}
      \State Announce $Tr_e$ as the earliest enabling transition
    \Else
      \State $process\_list \gets \{\}$
      \ForAll{$Tr \in q_c.outgoing\_transitions$}
        \If{$Tr \notin TrC$}
          \If{$p_i \not\in Tr$}
            \ForAll{$p_j \in Tr$}
              \State $process\_list.add(p_j)$
            \EndFor
          \Else
            \ForAll{$p_j \in Tr$ \textbf{s.t.} $lu_j < t_{Tr_e}$}
              \State $process\_list.add(p_j)$
            \EndFor
          \EndIf
        \EndIf
      \EndFor
      \ForAll{$p_j \in process\_list$}
        \State \textbf{send} $\langle Aggregate, t_{llc}, TrC, Tr_e, t_{Tr_e} \rangle$ to $p_j$
      \EndFor
    \EndIf
  \EndIf
\EndProcedure
\end{algorithmic}
\end{algorithm}

\noindent We now give an example of the entire algorithm in practice:
\begin{example}
  \begin{figure}[H]
    \scalebox{0.6}{
      \begin{tikzpicture}[scale=1]
        \draw [thick] (0,0) -- (9,0);
        \draw [thick,->,gray] (9,0) -- (12,0);
        \node[align=center] at (0.9,-0.7) {$\sigma_0 = \{\}$};
        \draw [thick,decorate,decoration={brace,amplitude=6pt,raise=0pt,mirror}] (-0.25,-0.15) -- (2.05,-0.15);
        \node[align=center] at (3.65,-0.7) {$\sigma_1 = \{a\}$};
        \draw [thick,decorate,decoration={brace,amplitude=6pt,raise=0pt,mirror}] (2.15,-0.15) -- (5.15,-0.15);
        \node[align=center] at (7.1,-0.7) {$\sigma_2 = \{a,b\}$};
        \draw [thick,decorate,decoration={brace,amplitude=6pt,raise=0pt,mirror}] (5.25,-0.15) -- (8.95,-0.15);
        \node[align=center,gray] at (10.65,-0.7) {$\sigma_3 = \{a,b,c\}$};
        \draw [thick,decorate,decoration={brace,amplitude=6pt,raise=0pt,mirror},gray] (9.05,-0.15) -- (12.25,-0.15);
        \node[align=center] at (2.1,1.3) {$t=2.1$};
        \draw [thick,->] (2.1,1) -- (2.1,0.15);
        \node[align=center] at (5.25,1.3) {$t=5.2$};
        \draw [thick,->] (5.2,1) -- (5.2,0.15);
        \node[align=center] at (9,1.3) {$t=9$};
        \draw [thick,->] (9,1) -- (9,0.15);
        \node[align=center,gray] at (11,0.8) {\textbf{Satisfaction}};
        \draw [line width=1.6pt,->,gray] (11,0.5) -- (9.3,0.15);
      \end{tikzpicture}
    }
    \scalebox{0.6}{
    	\begin{tikzpicture}[->,>=stealth',shorten >=1pt,auto,node distance=3.5cm,semithick]
        \tikzstyle{every state}=[fill=white,draw=black,text=black]
        \node[initial, state]   (S0) {$q_0$};
        \node[state]            (S1) [right of = S0] {$q_1$};
        \node[state, accepting] (ST) [above right of = S1] {$q_\top$};
        \node[state, accepting] (SF) [below right of = S1] {$q_\bot$};
        \path (S0) edge [out=30, in=150] node                         {$Tr_0 = a \wedge \lnot b$} (S1)
                   edge [out=-30, in =-150, below] node               {$Tr_1 = a \wedge \lnot c$} (S1)
                   edge [out=90, in=180] node                         {$Tr_2 = b \wedge c$} (ST)
              (S1) edge [out=10, in=-100, below] node [xshift=9mm]    {$Tr_3 = b \wedge c$} (ST)
                   edge [out=-10, in=100] node [yshift=-1mm]          {$Tr_4 = \lnot a \wedge \lnot b$} (SF)
                   edge [out=-90, in=180, below] node [xshift=-11mm]  {$Tr_5 = \lnot a \wedge \lnot c$} (SF);
      \end{tikzpicture}
    }
  \end{figure}
  Following the first scenario in Example~\ref{eg:computations}, and using the automaton in Fig.~\ref{eg:finalAutomaton}, at the beginning, all processes assume that the current automaton location is $q_0$. Say initially $p_1$ is the coordinator of $Tr_0$ and $p_3$ is the coordinator of $Tr_1$
  \\
  At $t = 2.1$, $a$ becomes $\mathit{true}$, and process $p_1$ which has access to $a$, realizes that $Tr_0$ can become enabled, but it doesn't know about $b$ yet since only $p_2$ has access to $b$, and hence, $p_1$ sends a Delegate message to $p_2$. The same happens for $p_3$ and $Tr_1$, resulting in $p_3$ sending a Delegate message to $p_1$. Upon receiving the message, $p_2$ realizes that $Tr_0$ was indeed enabled at $t = 2.1$, so it sends an Aggregate message to $p_3$. $p_1$ also realizes that $Tr_1$ was enabled at $t = 2.1$ after receiving $p_3$'s Delegate message, so it sends an Aggregate message to $p_2$. Since both $Tr_0$ and $Tr_1$ were enabled at the same time, whichever of $p_2$ or $p_3$ receives their Aggregate message first detects the automaton location change from $q_0$ to $q_1$ at $t = 2.1$. Say after detection, it picks $p_3$, $p_2$ and $p_1$ as the coordinators of $Tr_3$, $Tr_4$ and $Tr_5$, respectively. After getting chosen as $Tr_4$'s coordinator, $p_2$ immediately sends a Delegate message to $p_1$, because $\lnot b$ is $\mathit{true}$ at $t = 2.1$. Now $p_1$ is the coordinator for both $Tr_4$ and $Tr_5$.
  \\
  At $t = 5.2$, $b$ becomes $\mathit{true}$, but since $p_2$, which has access to $b$ is not a coordinator, it doesn't do anything and no message is sent in the system.
  \\
  At $t = 9$, $c$ becomes $\mathit{true}$, and hence, $p_3$ detects that there is a possibility for $Tr_3$ to become enabled at $t = 9$. Therefore, it sends a Delegate message to $p_2$. Upon reception, $pr_2$ detects that $Tr_3$ was indeed enabled at $t = 9$, so it sends an Aggregate message to $p_1$. After receiving the message, $p_1$ being the coordinator of both $Tr_4$ and $Tr_5$ comes to the conclusion that neither can be enabled before $t = 9$. Hence, it updates its automaton location to $q_\top$ and announces the satisfaction of the LTL property.
  \\\qedsymbol
\end{example}

\noindent \textbf{Note:} The algorithm can be slightly optimized by concatenating messages with the same destination in one message packet. For example, if $p_i$ is the coordinator of three transitions, and at time $t$, its literals become unsatisfied in the first two transitions' predicates, and if the next coordinator of both transitions is $p_j$, and also if at time $t$, process $p_i$ realizes that the third transition is enabled at a certain time, as a result, a single message can be sent to $p_j$, containing the Delegate messages for the first two transitions and the Aggregate message resulting from the third transitions being added to $p_i$'s $TrC$.

\section{Algorithm Proof}
\label{sect:proof}

This section provides the proof of the soundness and correctness of our algorithm.
\begin{itemize}
	\item Our algorithm is \textbf{sound}, meaning that if it announces the satisfaction/violation of the LTL property at time $t$, the property has actually been satisfied/violated at time $t$.
	\item Our algorithm is \textbf{complete}, meaning that if the LTL property gets satisfied/violated at time $t$, the algorithm will announce the property's satisfaction/violation at time $t$ (after a finite delay time from then).
\end{itemize}

\subsection{Soundness}

\noindent Assume, for the sake of contradiction, that: \textit{the algorithm announces the satisfaction/violation of the LTL property at time $t$ while in reality this claim isn't true.}
Based on this assumption, at one of the steps during the algorithm's execution, a wrong transition is taken, or the right transition is taken at the wrong time.
\begin{enumerate}
	\item In the case of the right transition being taken at the wrong time, if we assume $t_{detected}$ to be the time that the algorithm has detected the transition, and $t_{actual}$ to be the actual time that the transition should have been taken:
	$$t_{detected} \neq t_{actual} \Rightarrow (t_{detected} > t_{actual}) \vee (t_{detected} < t_{actual})$$
	\begin{itemize}
		\item 	In the case of $t_{detected} > t_{actual}$, given the fact that $t_{detected} = min(gpsr)$:
		$$t_{actual} < min(gpsr) \Rightarrow t_{actual} \not\in min(gpsr)$$
		As stated in the algorithm description, the initial value of $gpsr$ is $[t_{llc}, \infty)$ at the beginning of each step, and the only time it changes is when a process realizes that one or more of its literals are not satisfied at particular times, which are then removed  from $gpsr$. Given that $t_{actual}$ is in the time range of the current step, and assuming this is the first step of error, it is obvious that $t_{actual} \in [t_{llc}, \infty)$. That means that at some point, a coordinator process removed $t_{actual}$ from $gpsr$, because one or more of its literals were not satisfied at that time, which contradicts the fact that the transition was actually enabled at $t_{actual}$.
		\item In the case of $t_{detected} < t_{actual}$, since $t_{actual}$ is the earliest time the transition becomes enabled, therefore, the transition was disabled at $t_{detected}$.  Hence, one or more literals of a process $p_i$ were not satisfied at $t_{detected}$. We know that $t_{lu_i} \ge t_{detected}$, so since  $p_i$ was the coordinator at $t_{lu_i}$, and  ${t_{llc} \le t_{detected} \le t_{lu_i}}$, then $p_i$ must have removed $t_{detected}$ from $gpsr$, which contradicts $t_{detected}$ being picked by the algorithm as the transition's enabling time.
	\end{itemize}
	\item In the case of a wrong transition being taken, in reality another transition (which we name $Tr_{actual}$) must have been enabled earlier than the transition that the algorithm has detected (which we name $Tr_{detected}$). The process that detected $Tr_{detected}$ as the earliest enabling transition, must have had $Tr_{detected}$ as its $Tr_e$ at that time, and its $TrC$ must have contained all the outgoing transitions of $q_c$. If both $Tr_{actual}$ and $Tr_{detected}$ were added to its $TrC$ by an Aggregate message, we observe the sender of the Aggregate message; eventually we find a process that had one of the two transitions in its $TrC$ at some point in time, and later on, the other got added to its $TrC$. We previously proved that transitions' enabling time are never incorrectly detected, and given that whenever a transition gets added to $TrC$, its enabling time is compared to the process's $t_{Tr_e}$. Since $Tr_{actual}$ is enabled earlier than $Tr_{detected}$, it is impossible for $Tr_e$ to be $Tr_{detected}$ after the second transition is added to $TrC$. So this case also contradicts our initial assumption.
\end{enumerate}

\subsection{Completeness}

In the proof of soundness, we came to the conclusion that the earliest transitions and their enabling time never get detected incorrectly, and the automaton location changes detected by the algorithm are corresponding to what happens in reality. The only case left that could violate the algorithm's completeness, is that at some step, the algorithm halts and doesn't detect a location change, where in reality a transition (which we call $Tr_{taken}$) must have been enabled and taken at a particular time (which we call $t_{taken}$). We will proceed to analyze this step.

 Note that transitions only get added to processes' $TrC$ values and no $TrC$ ever has a transition removed from it. Therefore, for the algorithm to halt, there should exist a time  (which we call $t_{halt}$), when the  $TrC$ of all  processes stop changing. We define $TrCH_i$ to be the value of process $p_i$'s $TrC$ at $t_{halt}$.

We assume $TS = \{Tr_{\alpha_1}, Tr_{\alpha_2}, Tr_{\alpha_3}, \dots, Tr_{\alpha_k}\}$ to be the list of the $k$ outgoing transitions of $q_c$ sorted by their enabling time in increasing order. Therefore, $Tr_{\alpha_1} = Tr_{taken}$, and the transitions that never get enabled are placed at the end of the list. We now prove by induction, that at $t_{halt}$, there exists a process with all the transitions of $TS$ in its $TrC$ list, and hence, it can conclude the automaton location change. We define the induction hypothesis $P(m)$ as follows:

$$
\begin{tabular}{|p{0.01\textwidth}p{0.88\textwidth}p{0.01\textwidth}|}
  \hline
  &\ & \\
  &$P(m) =$ \textit{At $t_{halt}$, there exists a process containing all transitions $Tr_{\alpha_1}, Tr_{\alpha_2}, Tr_{\alpha_3}, \dots, Tr_{\alpha_m}$ in its $TrC$.}& \\
  &\ & \\
  &$\Rightarrow P(m) = \exists i\ \textbf{s.t.}\ \{Tr_{\alpha_1}, Tr_{\alpha_2}, Tr_{\alpha_3}, \dots, Tr_{\alpha_m}\} \subseteq TrCH_i$ &\\
  &\ & \\ \hline
\end{tabular}
$$

\noindent \textbf{Base Case:} $P(1)$ holds
$$\exists i \ \textbf{s.t.}\ Tr_{\alpha_1} \in TrCH_i$$
This is obvious, since $Tr_1 = Tr_{taken}$, and $Tr_{taken}$ eventually gets enabled, based on the proof of soundness, its enabling time will be detected by its coordinator and it will be added to the coordinator's $TrC$.\\

\noindent \textbf{Induction Step:} If $P(m)$ holds, then $P(m + 1)$ also holds
$$\exists i\ \textbf{s.t.}\ \{Tr_{\alpha_1}, Tr_{\alpha_2}, Tr_{\alpha_3}, \dots, Tr_{\alpha_m}\} \subseteq TrCH_i$$
$$\Rightarrow \exists j\ \textbf{s.t.}\ \{Tr_{\alpha_1}, Tr_{\alpha_2}, Tr_{\alpha_3}, \dots, Tr_{\alpha_{m + 1}}\} \subseteq TrCH_j$$

If $p_i$ is not associated with $Tr_{\alpha_{m + 1}}$'s predicate, the last time its $TrC$ value changes, it sends an Aggregate message containing $TrCH_i$ in its $TrC$, and with $Tr_{taken}$ as its $Tr_e$, to all processes associated with $Tr_{\alpha_{m + 1}}$'s predicate. One of those processes is $Tr_{\alpha_{m + 1}}$'s coordinator, which upon receiving the Aggregate message, realizes that $Tr_{\alpha_{m + 1}}$ was not enabled before $Tr_e$ which is $Tr_{taken}$. It then adds $Tr_{\alpha_{m + 1}}$ to its $TrC$ list, which previously contained transitions $Tr_{\alpha_1}$ through $Tr_{\alpha_m}$ from the received Aggregate message. \qedsymbol\\

If $p_i$ is associated with $Tr_{\alpha_{m + 1}}$'s predicate, the last time its $TrC$ value changes, it sends an Aggregate message containing $TrCH_i$ in its $TrC$, and with $Tr_{taken}$ as its $Tr_e$, to all processes associated with $Tr_{\alpha_{m + 1}}$'s predicate with a smaller $t_{lu}$ than $Tr_{taken}$. If one of those processes is $Tr_{\alpha_{m + 1}}$'s coordinator, similar to the previous case, the coordinator will end up with all transitions $Tr_{\alpha_1}$ through $Tr_{\alpha_{m + 1}}$ in its $TrC$. \qedsymbol

If none of those processes are $Tr_{\alpha_{m + 1}}$'s coordinator, that means they have passed Delegate messages along, until the coordinator became one of the processes that didn't previously receive the Aggregate message. That process's $t_{lu}$ is bigger than $Tr_{taken}$'s enabling time. That means the previous coordinator (which received the Aggregate message) now has all of its $t_{lu}$ values to be greater than $Tr_{taken}$'s enabling time, so upon receiving the Aggregate message, it can conclude that $Tr_{\alpha_{m + 1}}$ was not enabled before $Tr_{taken}$. It then adds $Tr_{\alpha_{m + 1}}$ to its $TrC$ list, resulting in it containing transitions $Tr_{\alpha_1}$ through $Tr_{\alpha_{m + 1}}$. \qedsymbol\\

\noindent \textbf{Conclusion:} $P(k)$ holds; \textit{At $t_{halt}$, there exists a process with all outgoing transitions of $q_c$ in its $TrC$ list, proving the algorithm's completeness.}

\section{Experimental Results}
\label{sect:experiments}
We have fully implemented our algorithm, and ran it in a simulated distributed environment. To evaluate our method, we have also implemented a centralized runtime verification algorithm, where each process sends any update in its local state to a central monitor.
The following properties are checked in our experiments:
\begin{itemize}
  \item $\phi_1$ is an extended version of our main example throughout the paper: \textit{``At some point, the leader arrives at the destination. It then stays there until all $k$ followers have also arrived"}. $\phi_1 = \lnot a \cup (a \cup (b_1 \wedge b_2 \wedge \dots \wedge b_k))$
  \item $\phi_2=$ \textit{``The leader is at the destination and stays there until all $k$ followers have also arrived"} $ = a \cup (b_1 \wedge b_2 \wedge \dots \wedge b_k)$
  \item $\phi_3=$ \textit{``Eventually, the leader and all the $k$ followers arrive at the destination"} $ = \lozenge(a \wedge b_1 \wedge b_2 \wedge \dots \wedge b_k)$
  \item And finally, the property used as the main example in \cite{menna}: $\phi_4 = \square(a \Rightarrow (b \cup c))$
\end{itemize}

\noindent The monitor automatons for the first three properties is depicted below in Figs~\ref{fig:phi1}, \ref{fig:phi2} and \ref{fig:phi3}:
\begin{figure}[H]
  \begin{center}
  \begin{tikzpicture}[->,>=stealth',shorten >=1pt,auto,node distance=3.5cm,semithick]
  \tikzstyle{every state}=[fill=white,draw=black,text=black]
  \node[initial, state]   (S0) {$q_0$};
  \node[state]            (S1) [right of = S0] {$q_1$};
  \node[state, accepting] (ST) [above right of = S1] {$q_\top$};
  \node[state, accepting] (SF) [below right of = S1] {$q_\bot$};
  \path (S0) edge [out=45, in=135] node {$a \wedge \lnot b_1$} (S1)
             edge node {$a \wedge \lnot b_2$} (S1)
             edge [out=-30, in=-150] node {$\dots$} (S1)
             edge [out=-60, in=-120, below] node {$a \wedge \lnot b_k$} (S1)
             edge [out=90, in=180, above] node [xshift=10mm, yshift=1mm] {$b_1 \wedge b_2 \wedge \dots \wedge b_k$} (ST)
        (S1) edge [above] node [rotate=45, scale=0.8] {$b_1 \wedge b_2 \wedge \dots \wedge b_k$} (ST)
             edge [out=0, in=90, above] node [rotate=-45] {$\lnot a \wedge \lnot b_1$} (SF)
             edge [out=-45, in=135, above] node [rotate=-45] {$\lnot a \wedge \lnot b_2$} (SF)
             edge [out=-75, in=-195, above] node [rotate=-45] {$\dots$} (SF)
             edge [out=-105, in=-165, below] node [rotate=-45] {$\lnot a \wedge \lnot b_k$} (SF);
  \end{tikzpicture}
  \end{center}
  \caption{The monitor automaton for the LTL property $\phi_1 = \lnot a \cup (a \cup (b_1 \wedge b_2 \wedge \dots \wedge b_k))$}
  \label{fig:phi1}
\end{figure}
\begin{figure}[H]
  \begin{center}
  \begin{tikzpicture}[->,>=stealth',shorten >=1pt,auto,node distance=3.5cm,semithick]
  \tikzstyle{every state}=[fill=white,draw=black,text=black]
  \node[initial, state]   (S0) {$q_0$};
  \node[state, accepting] (ST) [above right of = S0] {$q_\top$};
  \node[state, accepting] (SF) [below right of = S0] {$q_\bot$};
  \path (S0) edge [above] node [rotate=45, scale=0.8] {$b_1 \wedge b_2 \wedge \dots \wedge b_k$} (ST)
             edge [out=0, in=90, above] node [rotate=-45] {$\lnot a \wedge \lnot b_1$} (SF)
             edge [out=-45, in=135, above] node [rotate=-45] {$\lnot a \wedge \lnot b_2$} (SF)
             edge [out=-75, in=-195, above] node [rotate=-45] {$\dots$} (SF)
             edge [out=-105, in=-165, below] node [rotate=-45] {$\lnot a \wedge \lnot b_k$} (SF);
  \end{tikzpicture}
  \end{center}
  \caption{The monitor automaton for the LTL property $\phi_2 = a \cup (b_1 \wedge b_2 \wedge \dots \wedge b_k))$}
  \label{fig:phi2}
\end{figure}
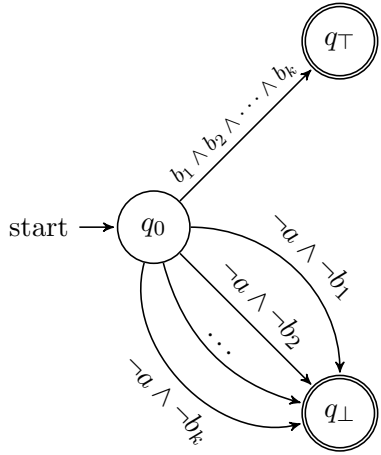
\begin{figure}[H]
  \begin{center}
  \begin{tikzpicture}[->,>=stealth',shorten >=1pt,auto,node distance=3.5cm,semithick]
  \tikzstyle{every state}=[fill=white,draw=black,text=black]
  \node[initial, state]   (S0) {$q_0$};
  \node[state, accepting] (ST) [right of = S0] {$q_\top$};
  \path (S0) edge node [scale=0.8] {$b_1 \wedge b_2 \wedge \dots \wedge b_k$} (ST);
  \end{tikzpicture}
  \end{center}
  \caption{The monitor automaton for the LTL property $\phi_3 = \lozenge(a \wedge b_1 \wedge b_2 \wedge \dots \wedge b_k)$}
  \label{fig:phi3}
\end{figure}

$\phi_1$, $\phi_2$ and $\phi_3$ are tested for $k=2$, $k=3$, $k=4$, $\dots$, $k=10$ (where $k$ is the number of followers). All properties may result in satisfaction or violation of the property, except for $\phi_3$, which can never be violated, and $\phi_4$, which can never be satisfied. We ran an experiment for each possible outcome in $\{\top, \bot, ?\}$. So, we had a total of $27$ experiments involving $\phi_1$ ($9$ for $\phi_1 \models \top$ for different values of $k \in \{2, 3, 4, \dots, 10\}$, $9$ for $\phi_1 \models \bot$, and $9$ for $\phi_1 \models\ ?$), $27$ experiments for $\phi_2$, $18$ experiments for $\phi_3$, and $2$ experiments for $\phi_4$, resulting in a total of 74 experiments. 600 randomized traces were generated for each of the 74 experiments. The traces were generated following a Poisson distribution over a range of 100 time units, 200 of the random traces were generated with the distribution's $\mu=10$, 200 with $\mu=100$, and 200 with $\mu=1000$. Message delays were assigned a random value in $[0, 2)$. We checked the value $\alpha$, defined as the ratio of messages sent if a centralized algorithm had verified the property, to the number of messages sent by our algorithm. This value is the \textit{improvement ratio} of our algorithm in terms of message efficiency.

Based on our experiments, the value of $\alpha$ was independent from $\mu$, or the properties truthfulness value, it only depended on the input itself. As observable in table~\ref{table:experiments}, properties of the first type have the least average improvement, since multiple transitions ($\lnot a \wedge \lnot b_1,\ \lnot a \wedge \lnot b_2,\ \dots$), all with a high chance of getting enabled, exit from the same automaton location, resulting in multiple Delegate messages being sent for each. This is while properties of the third type show the highest average improvement ratio, and in some cases, up to 60 fold fewer messages transferred, since only one transition has to be checked: $a \wedge b_1 \wedge b_2 \wedge \dots \wedge b_k$.

\begin{table}
\begin{tabular}{|l|l|l|l|}
\hline
\textbf{Property} & \textbf{Min. $\alpha$} & \textbf{Avg. $\alpha$} & \textbf{Max. $\alpha$} \\ \hline
$\lnot a \cup (a \cup (b_1 \wedge b_2 \wedge \dots \wedge b_{10}))$ & $0.479$ & $2.236$ & $16.831$ \\ \hline
$\lnot a \cup (a \cup (b_1 \wedge b_2 \wedge \dots \wedge b_9))$ & $0.585$ & $2.412$ & $10.905$ \\ \hline
$\lnot a \cup (a \cup (b_1 \wedge b_2 \wedge \dots \wedge b_8))$ & $0.668$ & $2.651$ & $13.028$ \\ \hline
$\lnot a \cup (a \cup (b_1 \wedge b_2 \wedge \dots \wedge b_7))$ & $0.896$ & $2.959$ & $11.318$ \\ \hline
$\lnot a \cup (a \cup (b_1 \wedge b_2 \wedge \dots \wedge b_6))$ & $1.132$  & $3.312$ & $15.331$ \\ \hline
$\lnot a \cup (a \cup (b_1 \wedge b_2 \wedge \dots \wedge b_5))$ & $1.549$ & $3.761$ & $14.638$ \\ \hline
$\lnot a \cup (a \cup (b_1 \wedge b_2 \wedge b_3 \wedge b_4))$ & $2.085$ & $4.310$ & $14.644$ \\ \hline
$\square(a \Rightarrow (b \cup c))$ & $2.059$ & $4.685$ & $6.138$ \\ \hline
$\lnot a \cup (a \cup (b_1 \wedge b_2 \wedge b_3))$ & $2.961$ & $4.983$ & $12.879$ \\ \hline
$\lnot a \cup (a \cup (b_1 \wedge b_2))$ & $4.753$ & $6.272$ & $12.514$ \\ \hline
$a \cup (b_1 \wedge b_2)$ & $8.024$ & $11.393$ & $15.639$ \\ \hline
$a \cup (b_1 \wedge b_2 \wedge b_3)$ & $7.120$ & $12.464$ & $19.659$ \\ \hline
$\lozenge(a \wedge b_1 \wedge b_2)$ & $7.883$ & $13.377$ & $20.423$ \\ \hline
$a \cup (b_1 \wedge b_2 \wedge b_3 \wedge b_4)$ & $5.685$ & $13.551$ & $24.625$ \\ \hline
$a \cup (b_1 \wedge b_2 \wedge \dots \wedge b_5)$  & $4.326$ & $14.481$ & $29.266$ \\ \hline
$\lozenge(a \wedge b_1 \wedge b_2 \wedge b_3)$ & $8.048$ & $15.321$ & $25.511$ \\ \hline
$a \cup (b_1 \wedge b_2 \wedge \dots \wedge b_6)$ & $3.092$  & $15.422$ & $33.645$ \\ \hline
$a \cup (b_1 \wedge b_2 \wedge \dots \wedge b_7)$ & $2.248$ & $16.490$ & $39.651$ \\ \hline
$\lozenge(a \wedge b_1 \wedge b_2 \wedge b_3 \wedge b_4)$ & $8.483$ & $17.367$ & $28.722$ \\ \hline
$a \cup (b_1 \wedge b_2 \wedge \dots \wedge b_8)$ & $1.641$ & $17.554$ & $44.043$ \\ \hline
$a \cup (b_1 \wedge b_2 \wedge \dots \wedge b_9)$ & $1.221$ & $18.619$ & $49.408$ \\ \hline
$\lozenge(a \wedge b_1 \wedge b_2 \wedge \dots \wedge b_5)$ & $8.385$ & $19.479$ & $35.186$ \\ \hline
$a \cup (b_1 \wedge b_2 \wedge \dots \wedge b_{10})$ & $0.928$  & $19.903$ & $55.395$ \\ \hline
$\lozenge(a \wedge b_1 \wedge b_2 \wedge \dots \wedge b_6)$ & $8.186$  & $21.738$ & $40.159$ \\ \hline
$\lozenge(a \wedge b_1 \wedge b_2 \wedge \dots \wedge b_7)$ & $8.689$ & $23.983$ & $46.116$ \\ \hline
$\lozenge(a \wedge b_1 \wedge b_2 \wedge \dots \wedge b_8)$ & $8.153$ & $26.456$ & $52.295$ \\ \hline
$\lozenge(a \wedge b_1 \wedge b_2 \wedge \dots \wedge b_9)$ & $8.801$ & $28.900$ & $55.640$ \\ \hline
$\lozenge(a \wedge b_1 \wedge b_2 \wedge \dots \wedge b_{10})$ & $8.373$ & $31.495$ & $60.571$ \\ \hline
\end{tabular}
\caption{Improvement in message efficiency for our different input properties}
\label{table:experiments}
\end{table}

\section{Related Work}
\label{sect:relatedWork}

A number of algorithms have been proposed for runtime verification in distributed systems. The authors in~\cite{centrMTL} and~\cite{centrMTLFreeze} provide  runtime verification algorithms for \textit{MTL} properties, and its extension with \textit{freeze quantifiers}. These algorithms  are both centralized, where all processes send their updates via messages to one or more central monitors that are responsible for checking the final verdict. Our work focuses on a distributed solution, which can decrease the number of required messages.

The algorithms introduced in \cite{sync,sync2,sync3} focus on runtime verification of LTL properties in \textit{synchronous systems}, where all the processes in the system experience local state updates at pre-defined fixed time points. These time points are also referred to as \textit{synchronous rounds}. The synchrony allows processes to communicate based on \textit{the $i$'th global state} in the system. We focus on asynchronous systems, where we cannot have any assumptions on synchrony of updates in different processes. 

Mostafa et al. in~\cite{menna} propose a distributed runtime verification algorithm for LTL properties in asynchronous systems. In the absence of a global clock, their algorithm cannot determine the total order of events, and hence, it can only provide us with \textit{possibilities} of satisfaction/violation of properties, based on the  different permutations of the total order of events. In this paper, we use the assumption of having a global clock, which is utilized to identify the order of events, and deterministically identify the satisfaction/violation of the monitored properties.

Some approaches define their own property definition languages, with syntaxes defined specifically for runtime verification. As an example, \cite{PT-DTL} defines a language called \textit{pt-DTL}, and~\cite{DTL} defines the extended version of that, called \textit{DTL}. Both of these languages are introduced for distributed systems, and their specifications involve processes' \textit{views of the system}, namely what each process knows about other processes. These languages, while intuitive, are not as expressive as LTL.

Part of our algorithm for detecting the satisfaction of a predicate is inspired from the methodology used in~\cite{slicing}. We added timing to that algorithm, and used it as a basis for our Delegate messages, which are used to detect the satisfaction of transitions' predicates.
The closest algorithm to our proposed approach is introduced by Falcone et al. in~\cite{falcone}. They propose a decentralized runtime verification algorithm for automata using global clocks. The algorithm enhances the given automaton with more states and transitions, and for each automaton location change,  it requires passing a message between all processes in a circular fashion. Obviously, their algorithm needs significantly more messages compared to our approach. Based on their reported results, their algorithm is  often less than $1.5$ times more efficient than a purely centralized algorithm, in terms of the number of messages passed. As mentioned in Section~\ref{sect:experiments}, we have achieved up to $60$ times less messages compared to a centralized solution.

\section{Conclusion}
\label{sect:conclusionFuture}

In this paper, we introduced a decentralized algorithm for runtime verification of LTL properties  (or any property describable by a DFA), with significant efficiency in terms of the number of messages passed in the system. Each process has a monitor that communicates with other monitors to identify the location changes in the given DFA. The main goal of monitors is to detect among the outgoing transitions of the current DFA location, the one with the earliest triggering predicate. We prove that our algorithm is both sound and complete. We have  implemented our algorithm, along with the centralized solution, and our experiments show that our decentralized solution can achieve up to 60 times fewer messages compared to the centralized one.

As for the future work, we plan to introduce time in the property specification, resulting in a timed automaton with clocks in its transition's predicates. We also plan to  propose an efficient runtime verification algorithm for these properties in a real-time distributed system.

\end{document}